\documentclass[a4paper,useAMS,usenatbib]{mn2e}

\usepackage[usenames]{color}
\usepackage{graphicx}
\usepackage{amsmath}
\usepackage{amssymb}
\usepackage{xspace}
\usepackage{ulem}
\usepackage{amsmath}
\usepackage{natbib}
\usepackage{amsbsy}

\newcommand{\diff}{{\rm{d}}}
\newcommand{\myvec}[1]{{\boldsymbol{#1}}}
\newcommand{\dm}{\ensuremath{{\rm{DM}}}}
\newcommand{\true}{\ensuremath{{\rm{true}}}}
\newcommand{\maxt}{\ensuremath{{\rm{max}}}}

\newcommand{\kin}{\ensuremath{{\rm{kin}}}}

\newcommand{\Msol}{\ensuremath{\text{M}_{\odot}}}
\newcommand{\Lsol}{{\rm L}_{\odot}}
\newcommand{\Mkpc}{\ensuremath{\text{M}_{1\text{kpc}}}\xspace}

\newcommand{\kms}{\ensuremath{{\rm km\,s}^{-1}}\xspace}
\newcommand{\Lmax}{\ensuremath{L_\text{max}}}

\begin{document}
\label{firstpage}

\title[Orbit-based dynamical models of the Sculptor dSph galaxy]{Orbit-based dynamical models of the Sculptor dSph galaxy}

\author[M. A. Breddels et al.]{
\parbox[t]{\textwidth}{
Maarten A. Breddels$^{1}$\thanks{E-mail:breddels@astro.rug.nl},
A. Helmi$^{1}$,
R.C.E. van den Bosch$^{2}$,
G. van de Ven$^{2}$,
and G. Battaglia$^{3}$
}
\\
\\
$^{1}$Kapteyn Astronomical Institute, University of Groningen, P.O. Box 800, 9700 AV Groningen, the Netherlands \\
$^2$Max Planck Institute for Astronomy, K\"onigstuhl 17, 69117
Heidelberg, Germany \\
$^3$\parbox[t]{\textwidth}{ European Organisation for Astronomical Research in the Southern
Hemisphere, K. Schwarzschild-Str. 2, 85748 Garching bei München,
Germany}\\
}

\maketitle

\begin{abstract}
  We have developed spherically symmetric dynamical models of dwarf
  spheroidal galaxies using Schwarzschild's orbit superposition
  method. This type of modelling yields constraints both on the total
  mass distribution (e.g. enclosed mass and scale radius) as well as
  on the orbital structure of the system (e.g. velocity
  anisotropy). This method is thus less prone to biases introduced by
  assumptions in comparison to the more commonly used Jeans modelling,
  and it allows us to derive the dark matter content in a robust
  way. Here we present our results for the Sculptor dwarf spheroidal
  galaxy, after testing our methods on mock data sets. We fit both the
  second and fourth velocity moment profile to break the
  mass-anisotropy degeneracy. {For an NFW dark matter halo
    profile}, we find that the mass of Sculptor within 1 kpc is $\Mkpc
  = (1.03 \pm 0.07 ) \times 10^{8}$ \Msol, and that its velocity
  anisotropy profile is tangentially biased and nearly constant with
  radius. The preferred
    concentration ($c \sim 15$) is low for its dark matter mass but
    consistent within the scatter found in N-body cosmological
    simulations. When we let the value of the central logarithmic slope $\alpha$ vary, we find that the
    best-fit model {has $\alpha = 0$}, although an NFW cusp or shallower is
    consistent at 1$\sigma$ confidence level. On the other hand, very
    cuspy density profiles with logarithmic central slopes $\alpha <
    -1.5$ are strongly disfavoured for Sculptor.
\end{abstract}

\begin{keywords}
galaxies: dwarf -- galaxies: kinematics and dynamics
\end{keywords}

\section{Introduction}

The existence of dark matter has been invoked to explain discrepancies in the observed kinematics of (systems of) galaxies. Especially in the last 30 years it has become a key ingredient of our current cosmological model, the $\Lambda$ cold dark matter paradigm (hereafter $\Lambda$CDM). N-body simulations have made clear predictions on how dark matter should be distributed in the Universe. Navarro, Frenk \& White (1996)  showed that simulated dark halos have a universal internal density distribution, now known as the NFW profile. Although there have been some revisions, the general form has remained, and the inner regions of simulated dark halos are found to be cusped with logarithmic slopes in the range $-1.2$ to $-0.75$ \citep{Navarro2010MNRAS.402...21N}. CDM simulations have also revealed the existence of a universal spin distribution and of relations between the characteristic parameters of a dark halo such as concentration and mass \citep[e.g][]{Bullock2001ApJ...555..240B}.

The predictions of the $\Lambda$CDM model may be tested using kinematic data. Cleaner tests are generally obtained using tracers located at large distances, i.e.\ in the regions that are dominated by the dark matter \citep[e.g][]{Romanowsky2003Sci...301.1696R,Battaglia2005MNRAS.364..433B,Battaglia2006MNRAS.370.1055B,Xue2008ApJ...684.1143X}. In these examples, a relatively accurate measurement of the mass contained within a given radius can be obtained, but constraints on the density profile depend on good knowledge of the spatial distribution of the tracers, which may be somewhat uncertain.  Another possibility is to use galaxies that are dark matter dominated at all radii, such as low surface brightness systems \citep{deBlok2010AdAst2010E...5D}.

An example of the latter class are the dwarf spheroidal (dSph) galaxies satellites of the Milky Way \citep{Mateo1998ARA}. These appear to be the most dark matter dominated galaxies with total dynamical mass to stellar light ratios in the order of 100-1000 $\Msol/\Lsol$ derived under the assumption of dynamical equilibrium \citep[e.g.][]{Wolf2010MNRAS.406.1220W}. The nearby dSph galaxies have the additional advantage that individual stars can be resolved, and their red giant branch (RGB) stars are bright enough to measure line-of-sight velocities with errors of a few km s$^{-1}$ \citep{Mateo1991AJ....102..914M}. The dynamical modelling of these objects is relatively simple since they are rather round, pressure supported and show little or no rotation. Their high dynamical mass-to-light ratios makes these systems ideal to study dark matter halos, especially their internal structure and to constrain their inner density profiles.

Most of the Milky Way dSph satellites have been modelled using the spherical Jeans equations \citep[e.g.][]{Kleyna2001ApJ...563L.115K,Battaglia2008ApJ...681L..13B,Strigari2008Natur.454.1096S,Lokas2009MNRAS.394L.102L,Walker2009ApJ...704.1274W}, while for more distant objects, such as the dSph satellites of M31, masses have been derived from the average velocity dispersion and projected mass estimators \citep{Kalirai2010ApJ...711..671K,Collins2010MNRAS.407.2411C}. In Jeans models one has to specify (i) the form of the light distribution, (ii) the density profile (or equivalently the gravitational potential) of the dark matter component, and (iii) velocity anisotropy of the stars. These characterise a given Jeans model, from which the second velocity moment projected along the line-of-sight can be computed. This is then compared to the measured line-of-sight velocity dispersion of the stars at different locations across the galaxy to establish the performance and characteristic parameters of the specific model.

Jeans modelling suffers from a number of limitations. Firstly the functional form of the velocity anisotropy has to be specified a priori while it is generally unknown. This is because precise measurements of the proper motions of stars in dSph are well beyond reach with current instrumentation. Also inherent to the method is the comparison between the moments of the model to those of the data which requires binning of the data and generally implies loss of information. It is also important to note that there is no guarantee that the resulting distribution function is non-negative everywhere, a requirement for it to be physical. Nonetheless, there have been interesting discoveries based the use of the Jeans equations and which are robust to assumptions of the underlying anisotropy profile. These include for example, the existence of a possible common mass scale
of dwarf spheroidals \citep[e.g.][]{Strigari2008Natur.454.1096S}, and the
tight constraints on the total mass within the half-light radius of these systems \citep{Walker2010ApJ...710..886W,Wolf2010MNRAS.406.1220W}.

Recently, \citet{An2009} demonstrated that if the tracer population is
supported by a spherical dark halo with a core or a cusp (less steep
than a singular isothermal sphere), then the central value of the logarithmic slope
$\gamma_0$ of the light profile and the central velocity anisotropy $\beta_0$ are
related as $\gamma_0 = 2 \beta_0$. This is valid if
$\sigma_r(0) > 0$, i.e. only if the stars are not dynamically cold in
this region. This would imply that the derived existence of a cusp or core at
the centre could merely be a consequence of the assumptions alone, if just the
second velocity moment is modelled using Jeans equations 
  \citep[see also][who show that a density slope-anisotropy inequality $\gamma > 2 \beta$
  holds at all radii, at least for a specific class of distribution
  functions for spherical systems]{Ciotti2010}. Thus care is required in interpreting the
  outcome of this type of models.

The above discussions shows clearly that there is a need to go beyond the modelling of the second moment using Jeans equations. For example, \citet{Lokas2002MNRAS.333..697L} proposed to use higher moments to constrain the internal dynamics of dSphs since the kurtosis profile depends mostly on anisotropy while the velocity dispersion depends both on mass and anisotropy \citet{Lokas2005}, hence this lifts some of the degeneracies.
Other possibilities would be to use parametrised phase-space distribution functions as pioneered by \citet{Kleyna2001ApJ...563L.115K,Kleyna2002,Wilkinson2002} \citep[see also][]{Amorisco2011}, or the made-to-measure technique \citep{syer-tremaine,Long2010}.

In this paper we take a different approach and use Schwarzschild modelling  \citep{Schwarzschild1979ApJ...232..236S} to probe the internal dynamics and characterise the dark matter content of the Sculptor dwarf spheroidal galaxy. The basic steps of the Schwarzschild method are to integrate a set of orbits in a given potential, calculate the predicted observables for each orbit, and then to weigh the orbits (with non-negative weights) to obtain a model that fits the observed data well in a $\chi^2$ sense. This approach guarantees that the distribution function (which is reflected in the orbit weights) is non-negative. This method was originally used by \citet{Schwarzschild1979ApJ...232..236S} to prove that a self consistent solution in dynamic equilibrium exists for a triaxial system, but was only implemented to reproduce the density distribution. The method was later extended to include kinematic constrains \citep{Richstone1984ApJ...286...27R,Pfenniger1984AA...141..171P}. Since then many codes have been 
developed \citep[e.g.][]{Richstone1984ApJ...286...27R,Rix1997ApJ...488..702R,vanderMarel1998ApJ...493..613V,Cretton1999ApJS..124..383C,Valluri2004ApJ...602...66V,vdBosch2008MNRAS.385..647V}. While first only the lowest moments of the line of sight velocity distribution (mean velocity and velocity dispersion) were fitted, better data have led to the inclusion of higher moments in the fits. While the use of moments allows one to use linear or quadratic programming to find the orbit weights, also likelihood methods using discrete data have been developed \citep[e.g.][]{Merrit1993AJ....106.2229M,Wu2006ApJ...643..210W,Chaname2008ApJ...682..841C}. A great advantage of Schwarzschild modelling is that it does not require the specification of the anisotropy profile, this is in fact an outcome of the model \citep[see also][for applications on the Fornax and Draco dwarf galaxies]{Jardel2012, Jardel2012Dra}.

Sculptor (Scl) is a dwarf spheroidal galaxy satellite of the Milky Way. It lies at high galactic latitude and is located at a heliocentric distance of 79 kpc. With an ellipticity of 0.32 (axis ratio is 0.68) it is not extremely flattened \citep{Irwin1995MNRAS.277.1354I}, allowing us to approximate and model Sculptor as a spherical object. Its luminosity is $L_V = 2.15 \times 10^6 L_\odot$ and one recent estimate of its dynamical mass is $2-3 \times 10^8 \Msol$ within 1.8 kpc \citep{Battaglia2008ApJ...681L..13B}. Its (stellar) mass distribution can be well fitted with a Plummer profile with scale radius $b=13.0$ arcmin \citep[$\simeq 0.3$ kpc,][]{BattagliaThesis}. Two large kinematic data sets have been compiled by \citet{Battaglia2008ApJ...681L..13B} and by \citet{Walker2009AJ....137.3100W}, leading to a total $\sim 2000$ member stars with radial velocity measurements with errors of $\sim2$ km/s. As we show below, the combination of these two data sets together with the Schwarzschild method allows us to 
constrain the dark matter distribution of Sculptor and its internal orbital structure.

This paper is organised as follows. In \S \ref{sec:model} we will describe the basic ingredients of Schwarzschild modelling, especially focusing on how it can be applied to dSph data. In \S \ref{sec:testing} we validate our model on a mock data set motivated by the current Sculptor data. In \S \ref{sec:scl} we apply the technique to Scl data, we present a brief discussion in \S \ref{sec:discussion}, and leave our conclusions to \S \ref{sec:conclusions}.

\section{Dynamical model}
\label{sec:model}

In this section we review some of the theory that provides the basis for our Schwarzschild method. We then describe how to generate models and focus later on how these can be fit to the observables.

\subsection{Generalities}
\label{sec:model:theory}
The phase-space structure of a galaxy can be specified by its distribution function (hereafter df) $f(\myvec{x}, \myvec{v})$, where $\myvec{x}$ and $\myvec{v}$ are the position and velocity coordinates respectively. The probability of finding a star in the volume $\diff {\myvec{x}}\diff {\myvec{v}}$ is given by $f(\myvec{x}, \myvec{v})\diff {\myvec{x}}\diff {\myvec{v}}$. All observables may be derived from knowledge of the df. For example the normalised surface density:
\begin{equation}
 \mu(x,y) = \int  \diff{}z \diff {\myvec{v}} f(\myvec{x}, \myvec{v}),
\end{equation}
where $z$ is the direction along the line-of-sight.

According to the (strong) \citet{Jeans1915MNRAS..76...70J} theorem, the df of a steady-state stellar system in which almost all orbits are regular, is a function of the isolating integrals of motion \citep[see also][]{BT}. Spherically symmetric systems (both in the tracer's density and the underlying potential) have only regular orbits and generally respect 4 integrals of motion, the energy and the 3 components of the angular momentum vector.
However, if the galaxy shows no rotation, due to symmetry, the df will depend only on the energy and the length of the angular momentum vector, i.e.\ $f(\myvec{x}, \myvec{v}) = f(E, L)$. Furthermore if the velocity distribution is isotropic, the df can only depend on energy  and $f(\myvec{x}, \myvec{v}) = f(E)$.

Most dSph galaxies are so distant that the only phase-space coordinates that may be measured currently are the projected stellar positions on the sky, and the line-of-sight velocities of (a subset of) its stars.
These can be used to derive the surface density $\mu_0(R)$ and the moments of the line-of-sight velocity distribution:

\begin{eqnarray}
 \mu_0(R) \!\!\!\! &=& \!\!\!\!\int \diff{}z \diff{}\myvec{v} f(E, L),\label{eq:df:surface_density}\\
 \mu_2(R) \!\! \!\!&=& \!\! \!\! \frac{1}{\mu_0(R)} \int  \diff{}z \diff{}\myvec{v} v^2_\parallel f(E, L), \label{eq:df:vdisp_los}\\
 \mu_4(R) \!\! \!\!&=& \!\! \!\! \frac{1}{\mu_0(R)} \int  \diff{}z \diff{}\myvec{v} v^4_\parallel f(E, L). \label{eq:df:vdisp_los4}
\end{eqnarray}
Here $R$ is the projected distance on the sky from the centre of the galaxy and $v_\parallel$ the velocity along the line-of-sight, after subtraction of the centre of mass mean motion.

The above equations suggest that through comparison to the observables it should be possible to derive the form of the df. In some cases, it may be better to parametrise the df and try to estimate its characteristic parameters by comparison to the data \citep{Wilkinson2002MNRAS.330..778W,Amorisco2011}.  However, in this work
we prefer to use a non-parametric approach such as the Schwarzschild method. This method uses orbits integrated in a specific gravitational potential as building blocks. From these, light and kinematical profiles may be derived and compared to observations through appropriate weighing of the orbits.

In the case of a dwarf galaxy embedded in a spherical dark matter halo, the gravitational potential can be characterised by a few parameters such as: i) the (enclosed) mass of the dark matter halo $M_\dm$, and ii) its scale parameter $r_\dm$. Due to the high dynamical mass-to-light ratios of dSphs, we do not expect the stellar mass to have a significant influence on the dynamics of the galaxy. We assume a fixed stellar mass-to-light ratio of $\Msol/\Lsol=1$ as in \citet{Walker2007ApJ...667L..53W}, and hence from the light distribution we may directly derive the gravitational potential associated to the stars. In the remainder of the paper we shall refer to properties related to the stellar mass and luminosity interchangeably.

Thus in practise, for a given set of parameters of the potential, we integrate orbits and match these to the observations by adjusting the orbital weights. We then repeat this exercise for other values of these parameters. This can be used to establish the values of the set of parameters which result in a better fit to the observables.

\subsection{From the model to the observables}
\label{sec:model:spherical}

Our Schwarzschild method is based on many of the ideas of \citet{Rix1997ApJ...488..702R} and \citet{vdBosch2008MNRAS.385..647V}. It is however, a new implementation that is optimised for spherical symmetry. Among other small improvements, our code can be run in parallel and is therefore significantly faster; furthermore for each orbit, we do not store the full line-of-sight velocity distribution but only its moments, which also reduces the computational load.

We now focus on how to generate the observables, namely the surface density and moments of the line-of-sight velocity distribution of the models and how to compare these to data.

For convenience we define $l=L/\Lmax$ the relative angular momentum (where $\Lmax$ is the angular momentum of a circular orbit of energy $E$), such that $l \in [0,1]$. This enables us to define a rectangular grid in energy and relative angular momentum. Since the Schwarzschild method is based on orbit integrations, the df may be seen as a sum of Dirac delta functions:
\begin{equation}
 f(E, L) = \sum_{i,j} \hat{f}_{i,j} \delta(E-E_i) \delta(L-l_j L_{\max,i}),
\end{equation}
where $\sum_{i,j} \hat{f}_{i,j} = 1$ and $\hat{f}_{i,j} \ge 0$.

To define the grid in energy and (relative) angular momentum we proceed as follows. For the energy we choose $N_E'$ radii between a minimum and maximum radius spaced logarithmically, and take the corresponding energy of a purely radial orbit. The minimum and maximum radii we consider are 0.033 kpc and 24.492 kpc, respectively. For each energy we choose $N_l'$ relative angular momenta spaced linearly between $0$ and $1$. All orbits are integrated starting from their apocentre.

We also define $N_R$ radial bins on the sky, defined by radii at the edges $R_k$ ($k=0...N_R$).  The borders are determined by the kinematic data, by requiring for instance that each bin contains a particular number of stars.

In general, it is convenient to work with the (normalised) mass in a given radial bin:
\begin{equation}
\begin{split}
\frac{\diff{}m_*(R)}{M_*} &= 2 \pi R \mu_0(R)\diff{}R.  \label{eq:diff_mass_distri}
\end{split}
\end{equation}
Thus the mass contributed by orbit of energy $E_i$ and relative angular momentum $l_j$ in the radial bin $k$ is:
\begin{equation}
\begin{split}
\frac{\Delta m_{*,i,j,k}}{M_*} &= \int_{R_k}^{R_{k+1}} 2 \pi R   \mu_{0,i,j}(R) \diff{}R. \label{eq:model_masses}
\end{split}
\end{equation}

In the Schwarzschild method this quantity is obtained by integrating the $i,j$ orbit and calculating the fractional time this orbit spends in radial bin $k$. Since we integrate the orbit with a fixed time step, this is simply equivalent to counting the number of times the orbit crosses bin $k$, divided by the number of time steps. To reflect the spherical symmetry, at each time step the position and velocities are rotated randomly $N_\text{rot} = 25$ times, as in \citet[][Eq.~2]{Rix1997ApJ...488..702R}. Each orbit is integrated for 100 orbital timescales $t_\text{orb}$, with $t_\text{orb} = 2\pi r_\text{a} / v_\text{circ}$, and where $r_\text{a}$ is the apocentre radius and $v_\text{circ}$ the circular velocity at $r_\text{a}$. Each orbit is stored at 1000 points (separated by a constant time step). Therefore the total mass  (contributed by all orbits) in bin $k$ is:
\begin{equation}
\begin{split}
\frac{\Delta m_{*k}}{M_*} &= \sum_{i=1}^{N_E'}\sum_{j=1}^{N_l'} g(E_i, L_j) \hat{f}_{i,j} L_\maxt \Delta E_i \Delta l_i \times \frac{\Delta  m_{*,i,j,k}}{M_*}\\
 &= \sum_{i=1}^{N_E'}\sum_{j=1}^{N_l'} c'_{i,j} \times \frac{\Delta  m_{*,i,j,k}}{M_*},
\end{split}
\end{equation}
where $g(E, L)$ is the density of states. The coefficients $c'_{i,j}$ are known as the orbital weights.

We may now proceed to calculate the light-weighted second and fourth moments of the line-of-sight velocity distribution in a given projected radial bin $k$ as:

\begin{eqnarray}
\mu_{2,k}\!\!\!\!\! &=& \!\!\!\!\!\frac{M_*}{\Delta m_{*k}} \sum_{i=1}^{N_E'}\sum_{j=1}^{N_l'} c'_{i,j} \int_{R_k}^{R_{k+1}} \!\!\!\!\!\! 2 \pi R \mu_{0,i,j}(R)\mu_{2,i,j}(R) \diff{}R,  \label{eq:model_mu2}\\
\mu_{4,k}\!\!\!\!\! &=& \!\!\!\!\!\frac{M_*}{\Delta m_{*k}} \sum_{i=1}^{N_E'}\sum_{j=1}^{N_l'} c'_{i,j} \int_{R_k}^{R_{k+1}} \!\!\!\!\!\! 2 \pi R \mu_{0,i,j}(R)\mu_{4,i,j}(R) \diff{}R,  \label{eq:model_mu4}
\end{eqnarray}
where $\mu_{2,i,j}(R)$ and $\mu_{4,i,j}(R)$ are the second and fourth moment respectively of orbit $i,j$.
The integral is also derived from the orbit integrations. However, instead of counting each time the orbit is found in bin $k$, we add the corresponding second moment in quadrature (and to the fourth power for the fourth moment) and at the end divide by the number of time steps. Note that the moments are linear in the orbital weights, which allows us to find a solution using quadratic programming, while for instance the kurtosis ($\gamma_2=\mu_4/\mu_2^2$) is not.

It is possible to consider the orbit weights ($c'_{i,j}$) as free parameters whose exact values will be determined through comparison to the observables. However this would imply that the number of orbits that are integrated to reproduce the observables is exactly equal to the number of free parameters that define the df. Decoupling these two sets of quantities is clearly desirable, see e.g. \citet{Cretton1999ApJS..124..383C}. This procedure is known as {\it dithering} and results in smoother density distributions while keeping the number of free parameters in the distribution function small.

While we may use $N_E' \times N_l'$ orbits to reproduce the observables, we choose only $N_E \times N_l = N_E' \times N_l' / (N_{d_E} \times N_{d_l})$ free parameters to characterise the distribution function, where we take $N_{d_E} \times N_{d_l} = 8\times 8 = 64$. The coefficients of the distribution function $c_{i,j}$ are related to the orbit weights ($c'_{i,j}$) as follows:
\begin{equation}
\begin{split}
   c'_{i,j} &= \frac{1}{N_{d_E} \times N_{d_l}} c_{i \backslash N_{d_E}, j \backslash N_{d_l}},
\end{split}
\end{equation}
where $\backslash$ indicates the integer part, e.g.\ $[i/N_{d_E}]$. Therefore $N_{d_E} \times N_{d_l}$ orbits share the same df coefficient. In practice, one can simply average the quantities obtained from the individual orbits. We choose $N_E = 20$ and $N_l = 8$, which results in $20\times8 = 160$ free parameters for the distribution function, but we integrate $20\times8\times8\times8=10250$ orbits.

To fit models to the data we generally use projected quantities (i.e. the observables). However, if one knows (or has derived) the df coefficients, it is also possible to make predictions for quantities that are not (yet) directly observable, such as the intrinsic (3d) density distribution or moments of the full velocity distribution.
For example, the mass contained in the (spherical) radial bin $m$ contributed by orbit $i,j$ is
\begin{equation}
\begin{split}
\frac{\Delta m_{*,3d,i,j,m}}{M_*} &= \int_{r_m}^{r_{m+1}} 4 \pi r^2 \nu_{*,i,j}(r) \diff{}r, \label{eq:diff_mass_distri_3d}
\end{split}
\end{equation}
where the integral is computed from the orbital integrations, and $\nu_{*,i,j}(r)$ is the radial density profile of orbit $i,j$. In practise we use
$N_r = 50$ (3d) radial bins, spaced linearly between $r_\text{min}=0$ kpc and $r_\text{max}=1.5$ kpc.
Similarly we also store the radial and tangential velocity dispersions in these bins.
Although we do not store the intrinsic properties beyond 1.5 kpc, this has no effect on the way the projected properties are determined. Note that the intrinsic properties are not used in any of the fitting routines but may be used for inferring for instance the intrinsic velocity anisotropy profile.

Orbits are integrated using the GNU Scientific Library (GSL) ordinary differential equation solver using an 8th order (Runge Kutta) Prince-Dormand method. We found that the energy is conserved to better than 0.1\%.

\subsection{Fitting procedure}
\subsubsection{Light distribution}

Our first requirement is for the model to fit the observed light distribution. We assume that this is known accurately. We require that the projected mass (or light) in each bin is matched within 1 per cent.  Given our assumption of a constant stellar mass-to-light ratio, we make no distinction between surface brightness and stellar mass surface density in what follows.  From the assumed brightness profile $\mu_*(R)$, we calculate:
\begin{equation}
 \frac{\Delta m_{*,\true, k}}{M_*} = \int_{R_{k}}^{R_{k+1}} 2 \pi R \mu_*(r)\diff{}R,
\end{equation}
and thus require for each projected radial bin $k$ that:
\begin{equation}
 \left| \frac{\Delta m_{*,\true, k}}{M_*} - \frac{\Delta m_{*,k}}{M_*} \right| \le 0.01. \label{eq:light}
\end{equation}

Note that the number of bins for the light does not have to equal the number of bins for the kinematics, in this work we choose 250 bins for fitting the light distribution.

\subsubsection{Kinematics}

To derive the line-of-sight velocity dispersion profile we calculate the second and fourth moment estimators of the line-of-sight velocity distribution $\hat{\mu}_{2,k}$ and $\hat{\mu}_{4,k}$ in bins containing at least $250$ stars. Assuming that the measurement errors are normally distributed, and all measurements and errors are independent and uncorrelated, we can obtain $\hat{\mu}_{2}$ of the population as follows. The expectation value of the second moment is
\begin{equation}
 E[m_{2}] = E\left[\frac{1}{N}\sum_i^{N} (v_i + \epsilon_i)^2\right] = \mu_{2} + s_2,
\end{equation}
where $\epsilon_i$ is the unknown noise of measurement $i$, which we assume is drawn from a normal distribution with dispersion $\sigma_i$ (i.e. this is the formal error of measurement $i$). Hence $s_2 = \left< \sigma_i^2 \right> = E\left[\frac{1}{N}\sum_i^N\epsilon_i^2\right] $ is the average of the estimated squared errors.  Here $\mu_{2}$ the true value of the second moment. Therefore, our best estimate for the second moment of the underlying population is:
\begin{equation}
 \hat{\mu}_{2} = \frac{1}{N}\sum_i^{N} (v_i + \epsilon_i)^2 - s_2. \label{eq:mom2}
\end{equation}
Similarly, the expectation value of the fourth moment
\begin{equation}
 E[m_{4}] = E\left[\frac{1}{N}\sum_i^{N} (v_i + \epsilon_i)^4\right] = \mu_{4} + 3s_2^2 + 6\mu_{2}s_2,
\end{equation}
where we have used that the fourth moment of a normal distribution is 3$\sigma^4$. Therefore our estimate for the fourth moment is:
\begin{equation}
 \hat{\mu}_{4} = \sum_i^{N} (v_i + \epsilon_i)^4 - 3s_2^2 + 6\mu_{2}s_2, \label{eq:mom4}
\end{equation}
where we have assumed $\mu_{2} \approx \hat{\mu}_{2}$.

The variance of the second moment $\text{var}(m_{2})$, can be determined using $\text{var}(x) = E[x^2] - (E[x])^2$, which yields
\begin{eqnarray}
 \text{var}(m_{2}) \!\!\!\!&=& \!\!\!\!\frac{1}{N} \left( \mu_4 - \mu_2^2 + 2 s_2^2 + 4\mu_2 s_2 \right) \label{eq:mom2_var}.
\end{eqnarray}
Although we formally need $\text{var}(\hat{\mu}_{2})$, we have found by testing with a Gaussian distribution, that for our purposes  $\text{var}(\hat{\mu}_{2}) \approx \text{var}(m_{2})$. For the variance of the fourth moment we find:
\begin{eqnarray}
\text{var}(m_{4}) \!\!\!\!&=& \!\!\!\! \mu_8 + 105 s_2^4 + 204 \mu_4 s_2^2+420 \mu_2 s_2^3 \nonumber \\
 &  + & 28\mu_6 s_2 - 9 s_2^4 \label{eq:mom4_var}
\end{eqnarray}
which require the $6^\text{th}$ and $8^\text{th}$ moments:
\begin{eqnarray}
E[m_{6}] &=& \mu_{6} + 15 \mu_4 s_2 + 45 \mu_2 s_2^2 + 15 s_2^3,\\
E[m_{8}] &=& \mu_{8} + 210 \mu_4 s_2^2 + 28 \mu_6 s_2 + 420 \mu2 s_2^3,
\end{eqnarray}
and again we use $\text{var}(\hat{\mu}_{4}) \approx \text{var}(m_{4})$.

The likelihood of the kinematic data given a model is:
\begin{equation}
 p(\text{kinematic data}|\text{model}) \propto e^{-\frac{1}{2} \chi_\text{kin}^2} \label{eq:p_kin}
\end{equation}
where
\begin{equation}
 \chi^2_{\kin} = \sum_k^{N_{\rm bins}} \frac{ \left( \hat{\mu}_{2, k} - \mu_{2,k} \right)^2}{\text{var}(\hat{\mu}_{2,k})} +  \sum_k^{N_{\rm bins}}  \frac{ \left( \hat{\mu}_{4,k} - \mu_{4,k} \right)^2}{\text{var}(\hat{\mu}_{4,k})}. \label{eq:chisq_kin}
\end{equation}
Here $\mu_{2,k}$ is given by Eq. (\ref{eq:model_mu2}), $\hat{\mu}_{2, k}$ is the estimate from the data for bin $k$ and similarly for the fourth moment\footnote{Here we have neglected correlations between the moments, although these may exist in practice.}.

\subsubsection{Finding a solution}
We need to find the $c_{i,j}$ that maximise the probability (Eq. \ref{eq:p_kin}) or minimise the $\chi_\text{kin}^2$, under the condition that all $c_{i,j}$ are positive (and sum up to unity) and the light distribution is reproduced to within 1 per cent. This problem can easily be solved by quadratic programming (QP), since the minimisation is quadratic in the df coefficients, and the constraints are linear. Note however that for this non-parametric problem, the parameter space is very large, and a solution will often yield an unrealistically spiky df.  To effectively reduce the parameter space and yield a smoother df, we add a regularisation constraint, in analogy to \citet{Cretton1999ApJS..124..383C} and \citet{vdBosch2008MNRAS.385..647V}, by including a penalty term to the total $\chi^2$. This term has the form:
\begin{subequations}
\begin{equation}
 \chi^2_\text{reg} = \chi^2_\text{reg,E} + \chi^2_\text{reg,L},
\end{equation}
\begin{equation}
 \chi^2_\text{reg,E} = \left( \lambda_E \sum_{j=0}^{N_L} \sum_{i=1}^{N_E-1} -\xi_{i-1} c_{i-1,j} + 2 \xi_{i}c_{i,j} - \xi_{i+1} c_{i+1,j} \right)^2,\label{eq:chisq_regE}
\end{equation}
\begin{equation}
 \chi^2_\text{reg,L} = \left( \lambda_L \sum_{j=1}^{N_L-1} \sum_{i=0}^{N_E} -\xi_{i} c_{i,j-1} + 2 \xi_{i}c_{i,j} - \xi_{i} c_{i,j+1} \right)^2,\label{eq:chisq_regL}
\end{equation}
\end{subequations}
where $\chi^2_\text{reg,E}$ and $\chi^2_\text{reg,L}$ are small for a smooth df. This smoothness requirement is implemented by demanding the second order derivatives of the df to be small, which we compute by taking second order finite differences (Eqs. \ref{eq:chisq_regE}-\ref{eq:chisq_regL}).

In our case we found $\lambda_L = \lambda_E/8$ to work well, and we calibrate $\lambda_E$ in the next section. The $\xi_{i}$ terms are the inverse of the (normalised) masses inside the radii defined by our energy grid (\S \ref{sec:model:spherical}) \citep[see also][Eq. 29]{vdBosch2008MNRAS.385..647V}. Since the regularisation term $\chi^2_\text{reg}$ is quadratic in the df coefficients, it can also be optimised using the QP.

The total $\chi^2$ now becomes:
\begin{equation}
 \chi^2 = \chi^2_\text{reg} + \chi^2_\kin, \label{eq:chisq}
\end{equation}
Minimising this equation, in combination with the linear constrains of the $c_{i,j}$ and the linear constraints on the light distribution (Eq. \ref{eq:light}) defines the problem for the QP.

\section{Testing the method}
\label{sec:testing}
\subsection{Plummer profile embedded in an NFW dark matter halo}
  \label{sec:testing:nfw}
\subsubsection{Mock Sculptor}

\begin{figure}
\centerline{\includegraphics[scale=0.42]{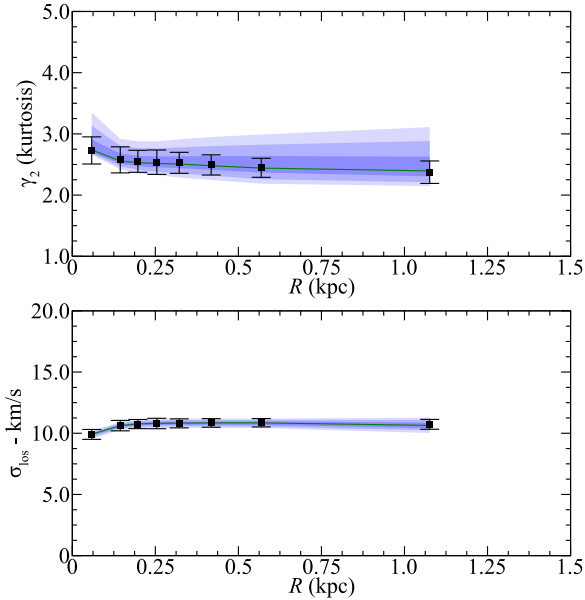}}
\caption{The line-of-sight velocity dispersion (bottom) and kurtosis (top) for mock Sculptor. {\bf Black symbols:}: Values for the moments in radial bins from the mock Sculptor data, with 1$\sigma$ error bars. {\bf Blue contours}: Recovered profiles from the models, where the regions correspond to the 68.3, 95.4 and 99.7 per cent confidence intervals.
\label{fig:mockscl-vlossigma}}
\end{figure}

\begin{figure}
\centerline{\includegraphics[scale=0.42]{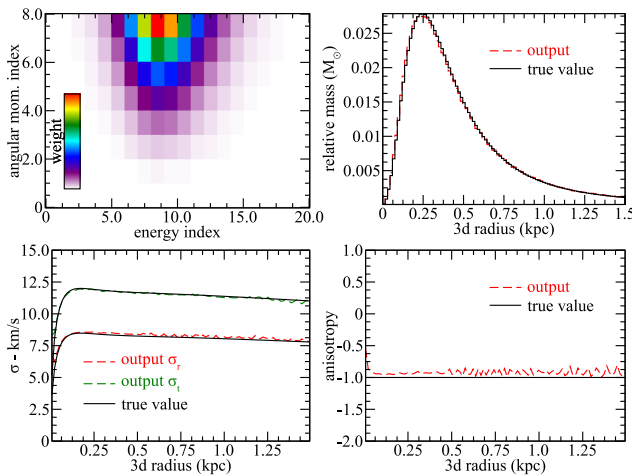}}
\caption{Result of a test of our Schwarzschild code on mock Sculptor. Here we have assumed knowledge of the df coefficients and recovered the intrinsic properties of the model.\label{fig:mockscl-known-results}}
\end{figure}

\begin{figure}
\centerline{\includegraphics[scale=0.42]{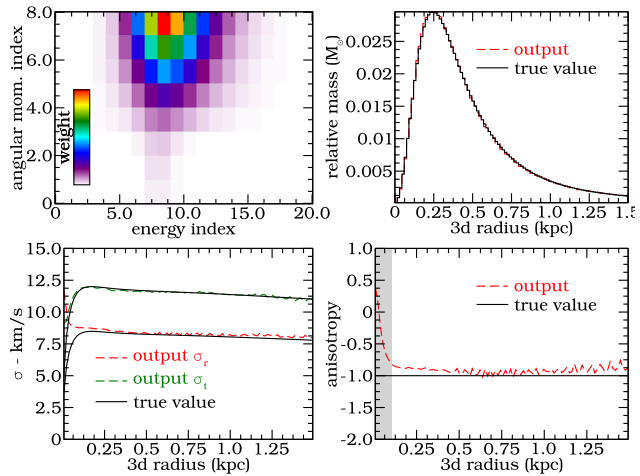}}
\caption{Result of the application of the Schwarzschild code on our mock Sculptor. The figure shows that the intrinsic structure is recovered through the QP when the underlying gravitational potential is known. The grey region in the lower right plot indicates where we cannot recover the anisotropy. \label{fig:mockscl-momentfitted-results}}
\end{figure}

We now create a mock galaxy that may be representative of Sculptor according to previously published dynamical models of this system \citep{Battaglia2008ApJ...681L..13B}. The goal is to test our method in the region of parameter space where we expect Sculptor to be. For the stellar component we choose a Plummer profile with total mass $M_* = 10^6$~\Msol ~and a scale radius $b = 0.3$ kpc. The stellar component is embedded in a spherical NFW dark matter halo with scale $r_s = 0.5$~kpc, and enclosed mass at 1 kpc of $M_\dm(< 1\,\text{kpc}) = 10^8$\Msol. The radial density profile for the NFW halo is of the form $\rho_{DM}(r)=\rho_0 (r/r_s)^{-1}(1+r/r_s)^{-2}$. We set the velocity anisotropy to be constant, $\beta = -1$. Recall that $\beta(r) = 1-\sigma_t^2(r)/\sigma_r^2(r)$ and $\sigma_t^2(r)$ (where  $\sigma_t^2=\sigma_\phi^2=\sigma_\theta^2$ for every $r$) and $\sigma_r^2(r)$ are the second moments of the intrinsic velocity distribution at radius $r$ in the tangential and radial directions respectively. Note 
that in this model, although the central velocity dispersion is null\footnote{which implies there is no conflict with the \citet{An2009} theorem.}, the line-of-sight velocity dispersion is finite, and has a value $\sigma_{los} = 7.71$~\kms. By assuming the df to be separable, i.e. $f(E,L) = f_E(E)f_L(L)$, we may compute it explicitly (numerically) as described in Appendix \ref{sec:df}.

As an extra check that our model galaxy is physical and stable, we have generated phase-space coordinates for $100~000$ stars from its df, and simulated it numerically using GADGET-2 \citep{Springel2005}. In this simulation the stars are represented as N-bodies and they are embedded in the static potential given by the dark halo of our mock Sculptor model. We found that,  even after $10$ Gyr of evolution, the density distribution, velocity dispersion profiles and the anisotropy match the initial values well.

To generate observations of our mock Sculptor we could draw a random sample of $\sim 2000$ stars from its distribution function. However, this has the disadvantage that many realisations would be required to test if the mean of the recovered quantities matches the known input values. Therefore, for the purposes of testing our modelling technique we prefer to compute the moments of the line-of-sight velocity distribution at different radii directly from the known distribution function,  as this is less susceptible to randomness. We add uncertainties in the moments and choose the location of the radial bins to match the Sculptor data set. Fig. \ref{fig:mockscl-vlossigma} shows the line of sight velocity dispersion profile and the kurtosis  derived in this way. Note however in the model fitting we use the second and fourth moments since these are linear in the df coefficients. We calculate
the uncertainties in the moments using Eqs.(\ref{eq:mom2_var}) and (\ref{eq:mom4_var}), assuming no measurement errors since these contribute only $\sim 1\%$ of the error budget for the typical measurement errors of 2~\kms and line of sight velocity dispersions of 10~\kms found in dSph.  Therefore the uncertainties in the moments are only due to the number of objects per bin. Here we
we choose to have 250 stars per bin, which gives a total 8 bins for a sample of 2000 objects.

We proceed to test our code in two steps. In the first instance our aim is to establish how well the method recovers the intrinsic properties of our mock galaxy if the df is known. Thus in this first test we use the known df to compute the df coefficients. These define the orbital weights which our Schwarzschild code uses to calculate the observables. The df coefficients are shown in the upper left panel of Fig. \ref{fig:mockscl-known-results}. The recovered (normalised) mass per intrinsic (3d) bin (Eq. \ref{eq:diff_mass_distri_3d}), is plotted in the top right panel of the same figure. The red dashed curve shows the output of the Schwarzschild code, while solid black corresponds to the true values. In the lower left panel we plot the velocity dispersions for the radial (red) and tangential (green) directions. The solid curves indicate the true values, whereas in dashed we showed the recovered dispersions. Here the ``true'' velocity dispersion has been calculated using the Jeans equations \citep[chapter 4]{
BT}. The lower right panel shows the true (solid black) and the recovered (dashed red) anisotropy as a function of radius. This exercise shows that given the correct weights we are indeed able to recover the known intrinsic properties of our mock galaxy.

The small deviations from the true values especially visible in the anisotropy profile are expected since the df coefficients only approximate the true df. These deviations can thus be removed by increasing the number of df coefficients. For example, if we double the number of coefficients in the energy and angular momentum directions, the small offset between the true and recovered anisotropy profiles disappears.  The increase in the resolution in the energy direction also leads to the elimination of the wiggles in the anisotropy profile. On the other hand, the turnover of the anisotropy profile seen at small radii is related to the sampling of orbits with the highest binding energy. Recall that we sample orbits from a minimum radius $r_{\rm min} \sim 0.03$ kpc, so that the highest binding energy radial orbit has its apocentre at $r_{\rm min}$.  The orbits that contribute to the region $r \lesssim r_{\rm min}$ are those which are very elongated with pericentres inside this radius and with large apocentres (
beyond $r_{\rm min}$), and the set of orbits with the highest binding energy but which have more angular momentum. These more circular orbits only contribute within a small range in radii, and hence the resulting velocity ellipsoid is radially biased. Clearly if we were to reduce $r_{\rm min}$, i.e. increase the sampling of orbits in the central regions, this will lead to a decrease in the radius at which the velocity anisotropy turns over. However, we deem this unnecessary as the amount of mass associated to this region is negligible, and this regime is in fact outside the reach of observations since we only have access to observables along the line-of-sight, and a star at small projected radius could be located at larger physical radii from the centre. Furthermore, the size of the currently available data sets is a strongly limiting factor (see next paragraph).

We now use the full Schwarzschild method, and solve for the df using QP. For the regularisation parameters we found $\lambda_E = 0.1$ to give good results. Fig. \ref{fig:mockscl-momentfitted-results} summarises our findings. The overall properties of the df are well recovered as well as the remaining characteristics (see Fig. \ref{fig:mockscl-known-results} for comparison). The anisotropy is recovered accurately except for  $r \lesssim$ 0.1 kpc. This is not due to sampling of highly-bound orbits discussed above, but is mostly driven by the small number of stars in this (3d) inner region. Running the same experiment with a larger data set (10\,000 and 50\,000 stars) we see the mismatch in the anisotropy to occur at smaller radii. In practise, this means that with the current data sets we are not sensitive to the anisotropy at $r \lesssim$ 0.1 kpc.

\subsubsection{Global halo parameter recovery}

\begin{figure}
\centerline{\includegraphics[scale=0.55]{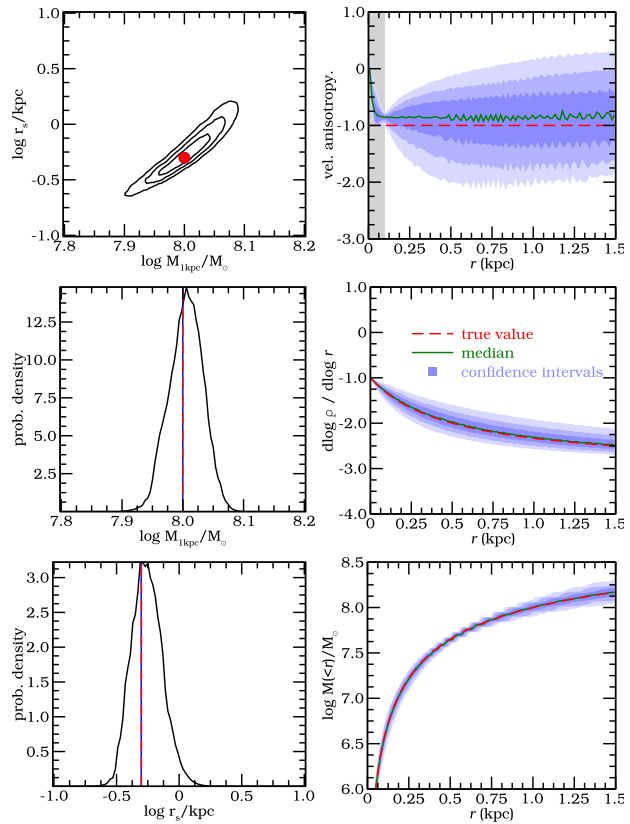}}
\caption{{\bf Left column:} Probability density functions (joint and marginalised) for mass and scale parameters of the NFW dark matter halo potential recovered for mock Sculptor model. Blue dot and blue lines (left column) indicate the maximum likelihood value (of the unmarginalised pdf), while the red dot and vertical dashed lines indicate the input values for the mock Sculptor model. The green solid line indicates the median value and the blue regions (or black contour lines in the top left panel) the 68.3, 95.4 and 99.7 per cent confidence intervals. {\bf Top right:} Recovered anisotropy profile. The grey region indicates where we cannot recover the anisotropy. {\bf Middle right:} Recovered logarithmic density slope (see text) for the dark matter. {\bf Bottom right:} Recovered enclosed mass profile. \label{fig:mockscl-pdf-mass-scale-and-others}}
\end{figure}

In the above tests we showed that the Schwarzschild code accurately recovers the df and therefore the kinematic properties of our mock dwarf galaxy. This test was done assuming that the (enclosed) mass within $1$ kpc of the NFW halo ($\Mkpc$) and its scale ($r_s$) were known. We now focus on how to estimate these parameters directly.

We proceed to calculate the probability of a model for a set of parameters values. In our case these parameters are $\Mkpc$ and $r_s$. However, instead of calculating this probability on a regular grid as done in e.g. \citet{Gebhardt2007ApJ...671.1321G} and \citet{vdBosch2008MNRAS.385..647V}, we use an adaptive method, similar to \citet{Gebhardt2009ApJ...700.1690G}. This first finds the probability density function (pdf) on a coarse grid and then determines where the pdf needs to be refined, and does so hierarchically. This allows us to obtain a relatively smooth pdf via the evaluation of a small number of models. For each set of model parameters we calculate the (relative) probability as $p \propto e^{-\frac{1}{2} \chi_\kin^2}$ (Eq. \ref{eq:chisq_kin}). This results in estimates of the best fit parameters, as well as in confidence intervals. We assume the prior on $\Mkpc$ to be uniform in $\log \Mkpc$ in the range $\log \Mkpc \in [7.6,8.2]$\footnote{Outside this interval the pdf is essentially zero.}  and 
the prior on $r_s$ uniform in $\log r_s$ in the range $\log r_s \in [-1, 1]$.

The pdf for the parameters \Mkpc and $r_s$ for our mock Sculptor model is shown in the top left panel of Fig. \ref{fig:mockscl-pdf-mass-scale-and-others}. The pdf is nicely centred on the input values $\Mkpc = 10^8\Msol$ and $r_s = 0.5 \approx 10^{-0.3}$ kpc. The maximum likelihood value (blue dot or lines) almost equals the input value, where the small deviation is caused by the discretisation of the pdf. Although the enclosed mass at 1 kpc is recovered both accurately and precisely (mean $\Mkpc = 1.02 \times 10^{8.00\pm0.03} \Msol$, corresponding to a 7\% uncertainty, or $\Mkpc = 1.02^{+0.075}_{-0.070} \times 10^{8} \Msol$), the scale radius is more poorly constrained (mean $r_s = 0.56\times10^{\pm 0.14}$ kpc, corresponding to a 37\% uncertainty, or $r_s = 0.56^{+0.21}_{-0.15}$ kpc). Note that the marginalised pdf for $\Mkpc$ and $r_s$ are somewhat asymmetric (a reflection of what is seen in the upper left panel of Fig.~\ref{fig:mockscl-pdf-mass-scale-and-others}), and this leads to slightly biased mean 
values for the parameters of the model.

Each Schwarzschild model (i.e. for a given \Mkpc and  $r_s$) results in a single anisotropy profile. To find the pdf of the velocity anisotropy profile one should integrate (marginalise) over all possible df coefficients \citep[as in][]{Magorrian2006MNRAS.373..425M}. However this is not always feasible due to the high dimensionality of the parameter space required to specify the
df ($N_E \times N_l = 160$ for this model). Instead we take the single anisotropy profile of each model, and calculate the probability density function for the anisotropy as a function of radius as follows:
\begin{equation}
p(\beta|r) = \int \diff{} \Mkpc \int \diff{} r_s p(\beta|r, \Mkpc, r_s)p(\Mkpc, r_s). \label{eq:anisotropy_at_r}
\end{equation}
We plot the median anisotropy as a function of radius in green in the top right panel of Fig.~\ref{fig:mockscl-pdf-mass-scale-and-others}, together with the 68.3, 95.4 and 99.7 percent confidence intervals in blue. Note however, that the anisotropy values at different radii are not independent. The input anisotropy is indicated by the red dashed line. The anisotropy seems to be reproduced quite accurately, except at small radii. Since our technique recovers nearly perfectly the input values of the model, the anisotropy profile found is essentially equivalent to that derived in Fig.~\ref{fig:mockscl-momentfitted-results}. The mismatch at small radii is explained in the previous section, and the apparent small uncertainty in the anisotropy in this region may be understood from the following argument. Using the Jeans equation, we may express the mass within a given radius as
$$GM(r)/r = \sigma_r^2 (\gamma - 2 \beta - \alpha),$$
where $\gamma = d\log \nu_*/d\log r$, $\beta$ is the anisotropy, and
$\alpha = d\log \sigma_r^2/d\log r$.  For any model without a black hole in the centre, the $lhs
\rightarrow 0$ as $r \rightarrow 0$. For a cored profile (as we have
assumed) $\gamma = 0$ in this limit. This implies that there is quite
a strong restriction on the behaviour of $\beta$ (and $\sigma_r$) at
small radii. The above equation implies that as $r \rightarrow  0$,
$$2 \sigma_r^2 - 2 \sigma_t^2 - r d\sigma_r^2/dr = 0,$$
and since $\sigma_r \rightarrow 0$ as $r \rightarrow
0$ to have a physical solution in a cuspy dark matter halo according
to \cite{An2009}, then this means that there is only one possible
$\sigma_t$ at $r = 0$, for any model, i.e. value of $M_{\rm DM}$ and $r_s$.

\begin{figure*}
\centerline{\includegraphics[scale=0.9]{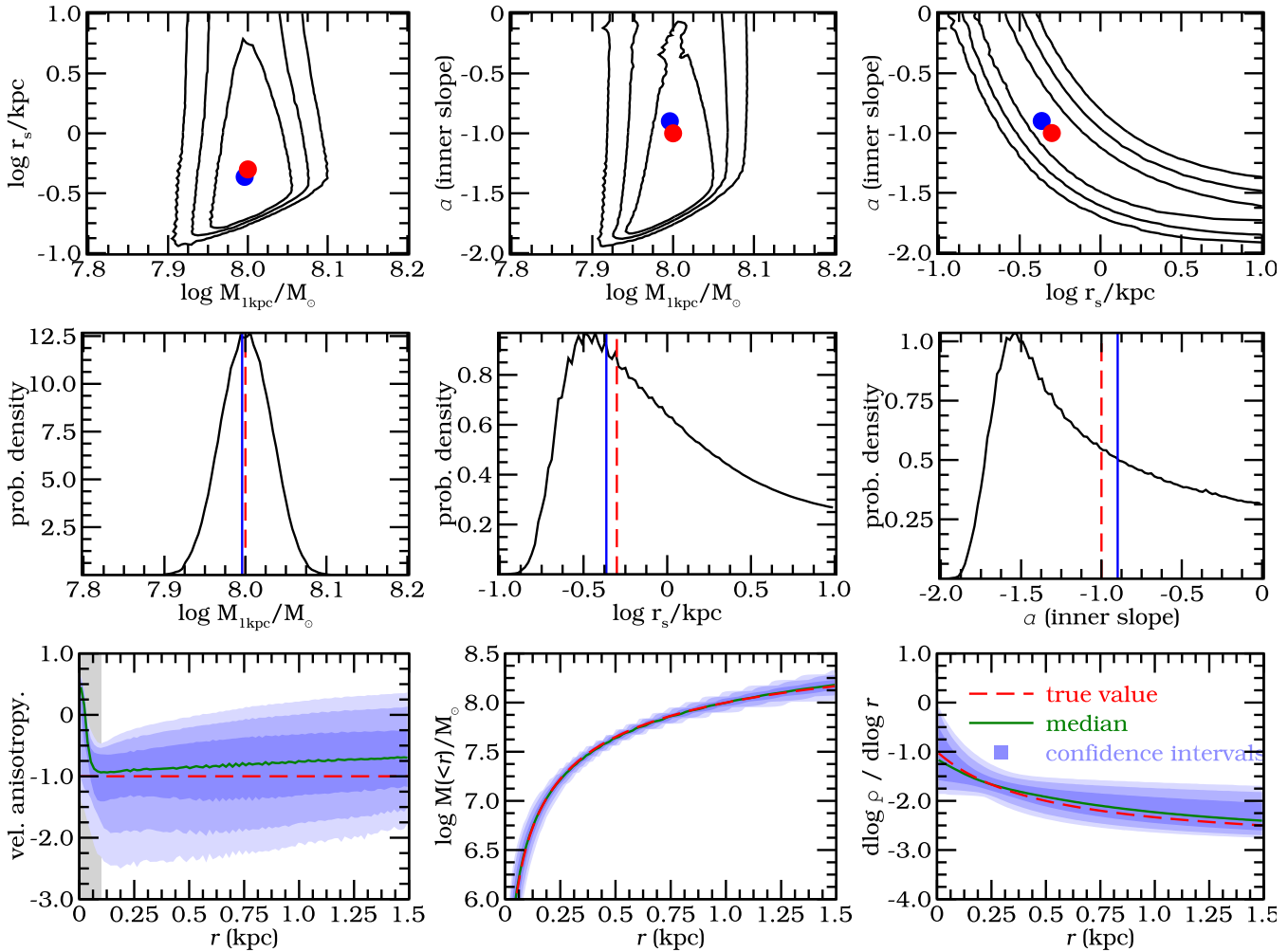}}
\caption{{\bf Top two rows:} Probability density functions (joint and marginalised) for mass, scale and inner slope parameters of the dark matter halo potential recovered for our mock Sculptor model. Blue dots (top row) and blue lines (middle row) indicate the maximum likelihood value (of the unmarginalised pdf), while the red dot and vertical dashed lines indicate the input values for the mock Sculptor model. The green solid line the median value and the blue regions (or black contour lines in the top row) the 68.3, 95.4 and 99.7 per cent confidence intervals. {\bf Bottom left:} Recovered anisotropy profile. The grey region indicates where we cannot recover the anisotropy. {\bf Bottom centre:} Recovered enclosed mass profile. {\bf Bottom right:} Recovered logarithmic density slope (see text) for the dark matter. } \label{fig:mockscl-pdf-mass-scale-slope-and-others}
\end{figure*}

In the central right panel of Fig.~\ref{fig:mockscl-pdf-mass-scale-and-others} we plot the logarithmic slope ($\eta$) of the dark matter density ($\rho_\dm$)  as a function of radius:
\begin{equation}
 \eta(r) = \frac{\diff{}\log \rho_\dm}{\diff \log r}.
\end{equation}
In the inner parts $\eta = -1$ and in the outer regions $\eta = -3$ due to our choice of the NFW profile. In the next section we also explore a different functional form for the halo density profile, which makes this plot more meaningful and useful for later comparison.

In the bottom right panel we plot the enclosed (dark matter) mass as a function of radius. The least uncertainty in the enclosed mass is at $r \approx 0.5-0.6$ kpc. This radius is close to the half light radius $r_{1/2} \approx 1.3 b \approx 0.4 $ kpc where \citet{Walker2009ApJ...704.1274W,Walker2010ApJ...710..886W} and \citet{Wolf2010MNRAS.406.1220W} find the enclosed mass to be most robustly determined and to be independent of anisotropy.

The line of sight velocity dispersion and the kurtosis profiles obtained from the models are shown as the blue contours in Fig.~\ref{fig:mockscl-vlossigma}. These have been computed in an analogous manner to the anisotropy profile, i.e. as in Eq.~(\ref{eq:anisotropy_at_r}). This figure shows that the resulting curves are in excellent agreement with the input profiles.

To gain further confidence in our methodology, we have also performed a similar set of tests for  different anisotropy profiles, while keeping the same stellar and dark matter density profiles. In one case the anisotropy varied from $\beta=-1$ in the centre to $\beta=+0.25$ at larger radii (i.e. from tangentially to radially biased). The other case we have tested has an anisotropy profile that changes from $\beta=0$ at the centre to $\beta=-1$ at larger radii (i.e. from radial to tangential anisotropy). Also in these cases all the quantities recovered are in excellent agreement with the input values, indicating that our methodology works well and is robust.

\subsection{Changing the dark matter halo density profile}
\label{sec:testing:general_halo}

In reality we will not know the actual density profile of the dark matter halo hosting a galaxy like Sculptor, and we would like to determine this from the data. A particularly interesting quantity is the inner slope of the density profile since this depends on the nature of the dark matter particles themselves, i.e.\ whether it is cold, warm or self-interacting \citep{Avila-Reese2001,Spergel2000}.

Therefore, in this section we use our mock Sculptor, which is embedded in an NFW profile, but
we assume a more general functional form to test the performance of our Schwarzschild method, i.e. we take:
\begin{equation}
 \rho_\dm(r) = \rho_0 \left(r/r_s\right)^\alpha \left( 1 + r/r_s\right)^{-(3+\alpha)},\label{eq:rho_general}
\end{equation}
such that for $\alpha = -1$ this reduces to the NFW case. For the orbit integration we need to know the potential (or rather the forces) generated by this density distribution. Since no general analytic expression exists for these general potentials, we have to solve Poisson's equation numerically. We do this using the FEM (Finite Element Method) method \citep[e.g.][]{pepper1992finite}. Our basis functions are Lagrange polynomials of degree $0$ to $3$ (cubic), which leads to a force field of order 2 (quadratic). We use a grid of 200 points in log radius, from $r=10^{-6}-10^4$ kpc. Testing this in the case of the NFW profile we find that the relative errors in the force in this range are $\sim 10^{-6}$.

We use our Schwarzschild code to find the best model that fits our mock Sculptor data, now with an additional unknown parameter $\alpha$, assuming a uniform prior in the range $\in [-2,0]$ ($\alpha > 0$ corresponds to a central hole in the dark matter distribution, which we do not consider). The results are given in Fig.~\ref{fig:mockscl-pdf-mass-scale-slope-and-others}. The top row in
shows the joint pdfs, marginalised over the remaining parameter. The middle row shows the pdfs of the single parameters, marginalised over the other two parameters. The blue dots and blue lines indicate the maximum likelihood value (of the unmarginalised pdf). In the bottom row the recovered anisotropy, mass and density profile are shown.

In general, all quantities are recovered quite well. However, the pdf of $\alpha$ versus $\log r_s$ shows an important degeneracy between these parameters, indicating that it is hard to determine either of these quantities reliably from our mock data set. The maximum likelihood (the blue dot) is slightly offset from the input value (red dot), which may indicate small systematic errors due to for instance the discretisation of the distribution function. However, note that since this systematic offset is in the direction of the degeneracy, the systematic error is small compared to the statistical uncertainty and therefore we do no consider this to be a problem for data sets of this size and quality. This analysis suggests that the current data is not sufficient to provide a good estimate of the inner slope for these models. The limitation lies in the number of stars with spectroscopic measurements (which in the case tested here is 2000) and/or their spatial distribution.

\section{Application to the Sculptor dSph galaxy}
\label{sec:scl}

\begin{figure} 
\includegraphics[scale=0.5]{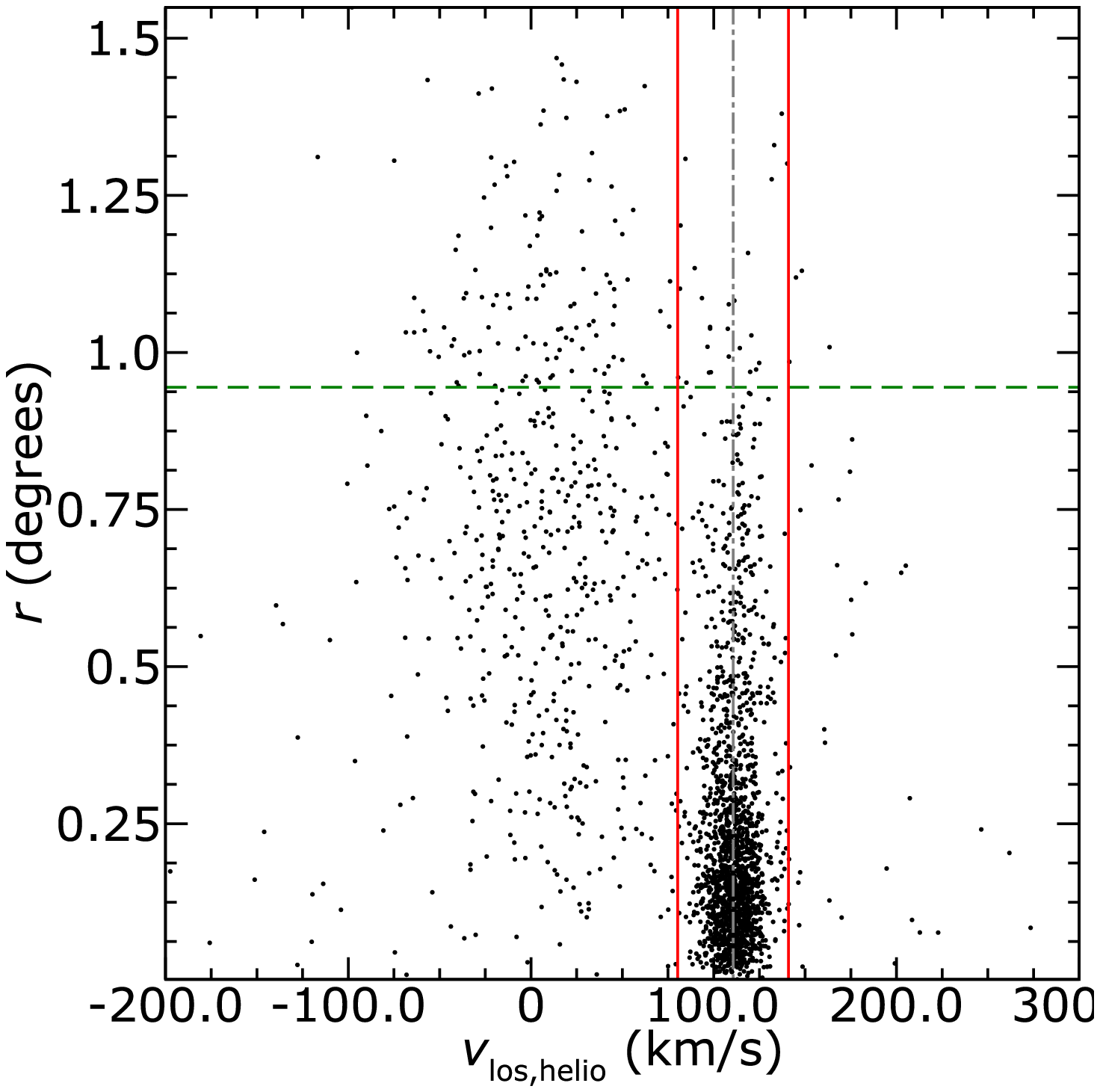} \includegraphics[scale=0.5]{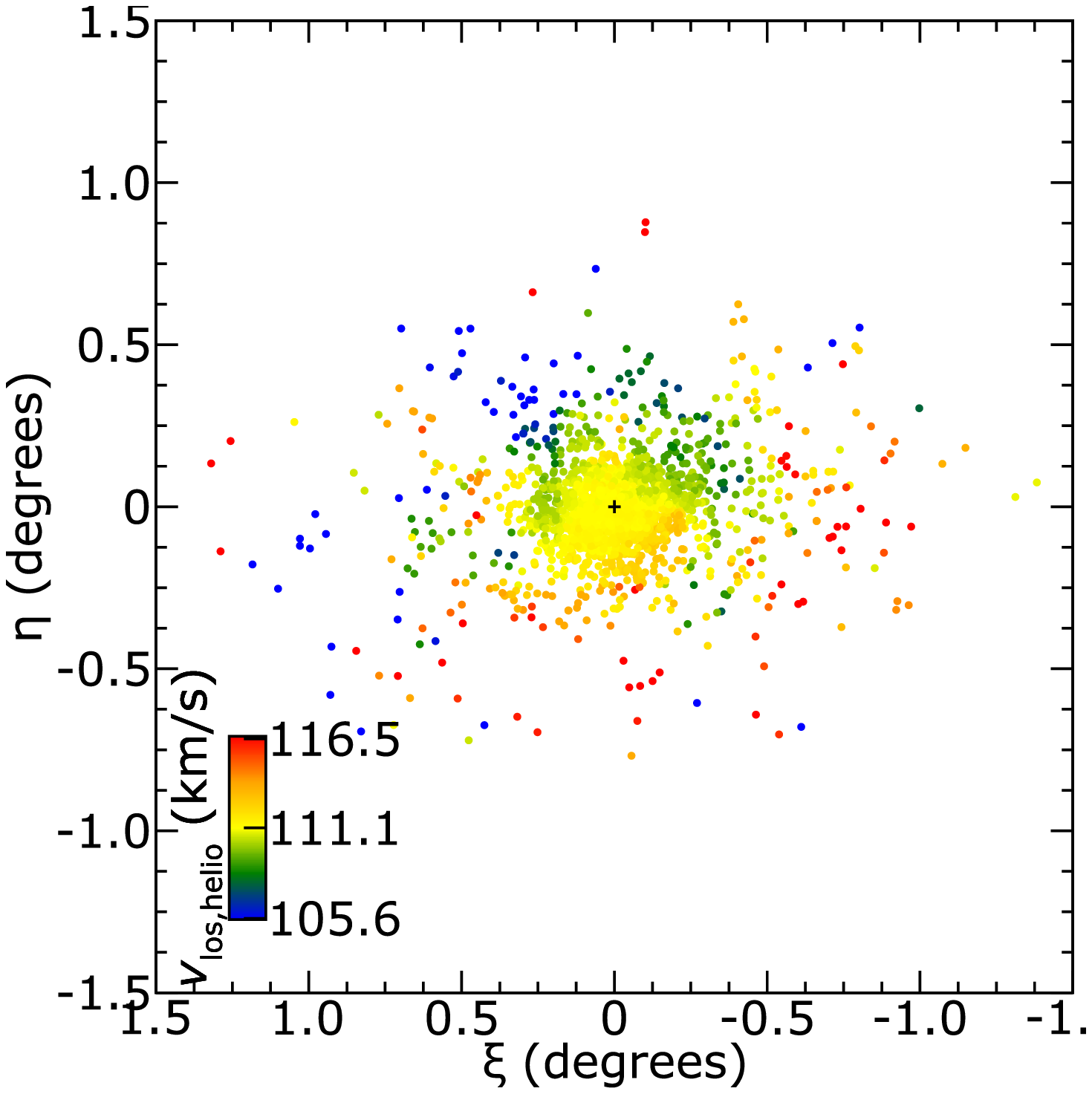}
\caption{{\bf Top:} Line of sight velocities for Sculptor versus radius. We only use the stars at radii $r <$ 3400 arcsec. The grey line indicates the systemic radial velocity for Sculptor and red lines $\pm3\sigma$ the mean velocity dispersion. {\bf Bottom:} Heliocentric line-of-sight velocities from the combined data set of \citet{Battaglia2008ApJ...681L..13B} and \citet{Walker2009AJ....137.3100W}. The velocities have been smoothed by taking the median in cells of 0.2 degrees on a side. \label{fig:scl-vlos}} \end{figure}

\subsection{Data and extracted velocity moments}
\label{sec:scl:data}

We use the line-of-sight velocities of
\citet[][1073 stars, hereafter B08]{Battaglia2008ApJ...681L..13B}and
\citet[][1541 stars, hereafter W09]{Walker2009AJ....137.3100W}\footnote{Although
\citet{boeboe} have reported a
systematic velocity offset of 1.5 km/s in the dataset of W09 compared to B08's, we here perform
no correction. The various tests we have done show that this offset has no visible effect on the results.}.
In the case of duplicates (stars in common in the datasets) we average
the line-of-sight velocities and the errors (in quadrature). Two
observations are considered to be from the same star when the
astrometry agrees within 1 arcsec, and a velocity difference less than
3$\epsilon$, where $\epsilon$ is the average velocity error. Inspection of the relative distances between
stars in the datasets shows that this criterion is optimal to sieve duplicates.
This procedure led to the identification of 308
duplicates, roughly 11\% of the combined dataset.

To create a velocity dispersion profile, we first need to convert the
measurements of the heliocentric line-of-sight velocities into
line-of-sight velocities that take into account the space motion of
Sculptor. We provide below a brief summary of this procedure and refer
the reader to Appendix \ref{sec:app:center_of_mass} for more details.

The heliocentric line-of-sight velocities of Sculptor's stars are shown in Fig.~\ref{fig:scl-vlos}. As can be seen from this figure, there appears to be a velocity gradient along the major axis (see also Fig.~1 of B08). The presence of such a gradient could be due to intrinsic rotation in
Sculptor, as suggested by B08. On the other hand it is also possible that the gradient is a result of the projection of the proper motion of the centre of mass of Sculptor (or a mix of both), in which case it can be used to infer its space velocity \citep{Walker2008ApJ}.
In absence of independent and direct measurements of the proper motion of Sculptor, it remains debatable what the source of the gradient is. For simplicity, here we assume that Sculptor does not rotate and we derive the velocity of the centre of mass of Sculptor from the line-of-sight measurements in Appendix \ref{sec:app:center_of_mass}. We note that in practice, our procedure simply removes the gradient, which one might say is equivalent to having removed (solid body) rotation.

\subsubsection{Velocity dispersion profile}
\label{sec:scl:data:vdisp}

For our dynamic modelling, we need to calculate the velocity dispersion profile of Sculptor in radial bins. To this end, we initially make a rough selection of the likely members of Sculptor, and then perform a more thorough analysis including the effects of Milky Way contaminants. In the first step, we take the systemic heliocentric radial velocity ($v_\text{Scl,sys,helio} = 110.6$ km s$^{-1}$)  and the mean velocity dispersion ($\sigma_\text{Scl} = 10.1$ km s$^{-1}$) from B08. We require that the member stars are within $3\sigma$ of the systemic velocity of Sculptor, as indicated by the red solid lines in the right panel of Fig.~\ref{fig:scl-vlos}. Furthermore we also require that they are located within $r~<~3400$~arcsec ($\sim 0.94$~deg, $1.3$ kpc), indicated by the green dashed line in the same panel. We add this requirement since we are not confident that outside this radius a reliable velocity dispersion can be measured due to the low number density of (probable) Sculptor members compared to Milky Way 
stars. An improved method for  discriminating Milky Way contaminants based on surface gravity in the data set of B08 has been developed by \citep{Battaglia2011AA}, see also \citet{Walker2009AJ....137.3109W}.

We then define radial bins such that each but the last bin contains at least 250 stars that match these criteria.
From the total of unique (i.e. non-duplicates) 2306 stars, 1695 match the above two criteria, resulting in 7 radial bins, where the last one contains 195 stars.

After we defined our bins to include at least 250 probable members, we remove the requirement of being within $3\sigma$ of the systemic velocity of Sculptor. We now only require $r<3400$ arcsec ($=1.3$ kpc), so all stars below the green dashed line in top panel of Fig.~\ref{fig:scl-vlos} are considered for calculating the velocity dispersions (2153 stars). We now use a model for the velocity distribution of the foreground contamination and of Sculptor itself, which then allows us to calculate the most likely velocity dispersion in each radial bin.

Following B08 and W09 we model the velocity distribution in a radial bin as a sum of Gaussians. The velocity distribution of Sculptor itself is modelled as a single Gaussian, while that of the Milky Way is modelled as a sum of two Gaussians\footnote{This gives a good fit to the Besan\c{c}on model in this region of the sky and for stars with colours and magnitudes in the observed range \citep{Robin2003A&A...409..523R}.}, following B08. Then the probability of the velocity dispersion of Sculptor in radial bin $j$ with data $D_j$ is:
\begin{equation}
\begin{split}\label{eq:prob_sigma}
 p(\sigma_j|D_j) &= \frac{p(D_j|\sigma_j)p(\sigma_j)}{p(D_j)} =
 	\prod_i^{N_j} \frac{p(D_{j,i}|\sigma_j)p(\sigma_j)}{p(D_{j,i})}\\
 		&= \prod_i^{N_j} \frac{p(R_{j,i},v_{j,i}|\sigma_j)p(\sigma_j)}{p(R_{j,i},v_{j,i})}\\
 		&\propto p(\sigma_j) \prod_i^{N_j} \left( p(R_{j,i},v_{j,i},m|\sigma_j) \right. \\
		&~~~~~~~~~~~~~\left. +p(R_{j,i},v_{j,i},\lnot m|\sigma_j) \right),
\end{split}
\end{equation}
where $R_{j,i}$ and $v_{j,i}$ are the radius and velocity of the $i^\text{th}$ star in the $j^\text{th}$ bin,  $N_j$ is the number of stars in bin $j$, $p(\sigma_j)$ is the prior, which we take flat between the range $0~\le \sigma_j~\le~30$~km~s$^{-1}$ and $m$ and $\lnot m$ indicate the Boolean value of being a member star of Sculptor or not. The proportionality can be used since the denominator is a normalisation constant. The first terms in the last line of Eq. (\ref{eq:prob_sigma}) can be expanded further (for each $j$):
\begin{equation}
\begin{split}
p(R_i,v_i,m|\sigma)	&= p(R_i,v_i|m,\sigma)p(m|\sigma)\\
			&= p(R_i,v_i|m,\sigma)p(m)\\
			&= p(R_i|m)p(v_i|m,\sigma)p(m)\label{eq:scl_pRvm}
\end{split}
\end{equation}
We take the prior on membership, to be equal $p(m) = p(\lnot m) = \frac{1}{2}$. Using the model of Sculptor as described above, $p(R_i|m) = \mu_\text{Scl}(R_i)$, the normalised surface density and $p(v_i|m,\sigma)$ is a Gaussian convolved with the individual measurement errors on $v_i$.

The second term in Eq. (\ref{eq:prob_sigma}) can similarly be derived
by replacing $m$ with $\lnot m$ in Eq. (\ref{eq:scl_pRvm}),
$p(R_i|\lnot m) = \mu_\text{MW} $ is the density of the Milky Way
foreground. Since the normalisation is not important, we only need to
know the ratio $\mu_\text{MW}/\mu_\text{Scl}(R_i)$ in each bin\footnote{Although the sampling of
  B08 and W09 is different, we
  have found in tests that this has no influence on our results.} . This
can be estimated by the ratio of stars outside the $3\sigma$ and
inside the $3\sigma$ velocity dispersion.  Furthermore, if there
is any bias in the sampling of the kinematic data (which usually is
the case), it will affect both the Sculptor data and the foreground
data in equal ways, and will cancel out in the ratio.  The term
$p(v_i|\lnot m,\sigma)$ is the weighted sum of two Gaussians as
described in \citet{Battaglia2008ApJ...681L..13B}.

\begin{figure}
\centerline{\includegraphics[scale=0.4]{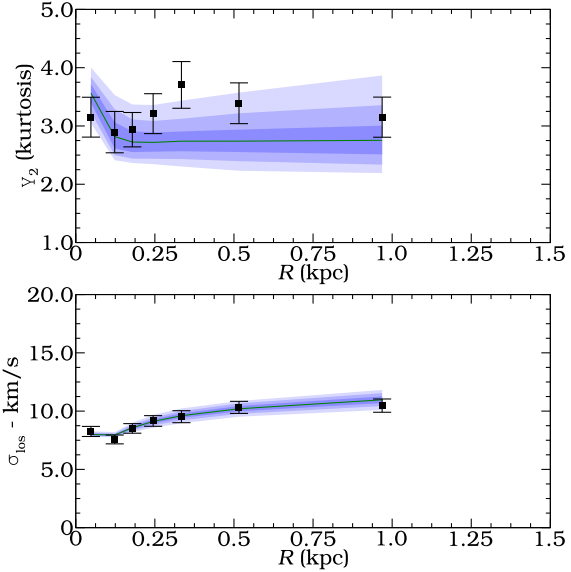}}
\caption{Line-of-sight velocity dispersion profile obtained in radial bins for the data by \citet{Battaglia2008ApJ...681L..13B} and \citet{Walker2009AJ....137.3100W} of Sculptor, taking into account the foreground contamination by the Milky Way. The dashed curve corresponds to the (pdf weighted) median line-of-sight velocity dispersion profile from the Schwarzschild models presented in Sec. 4.2, while the contours indicate the 1, 2, and 3$\sigma$ uncertainties around this curve. The last bin extends to 1.3 kpc. \label{fig:scl-velocity-dispersion-profile}}
\end{figure}

For each radial bin we find the maximum likelihood value for the
velocity dispersion. After this, we perform a $3\sigma$ clipping
around the mean, and estimate the second and fourth moments for the
remaining stars using Eqs.~(\ref{eq:mom2}) and (\ref{eq:mom4}). The
errors are computed from Eqs.~(\ref{eq:mom2_var}) and
(\ref{eq:mom4_var}).  The final sample contains 1696 member
stars. Fig.~\ref{fig:scl-velocity-dispersion-profile} shows the
resulting velocity dispersion profile and the kurtosis
($\hat{\mu}_4/\hat{\mu}_2^2$). The line-of-sight velocity dispersion
is well-constrained, it is relatively flat although it appears to be
slightly rising with radius. The kurtosis has larger error bars, and this implies that
additional modelling is required to establish in a robust statistical way what the shape of the velocity ellipsoid is
\citep{gerhard1993}.

\subsection{Schwarzschild method applied to Sculptor}
\label{sec:scl:apply}
\begin{figure}
\centerline{\includegraphics[scale=0.6]{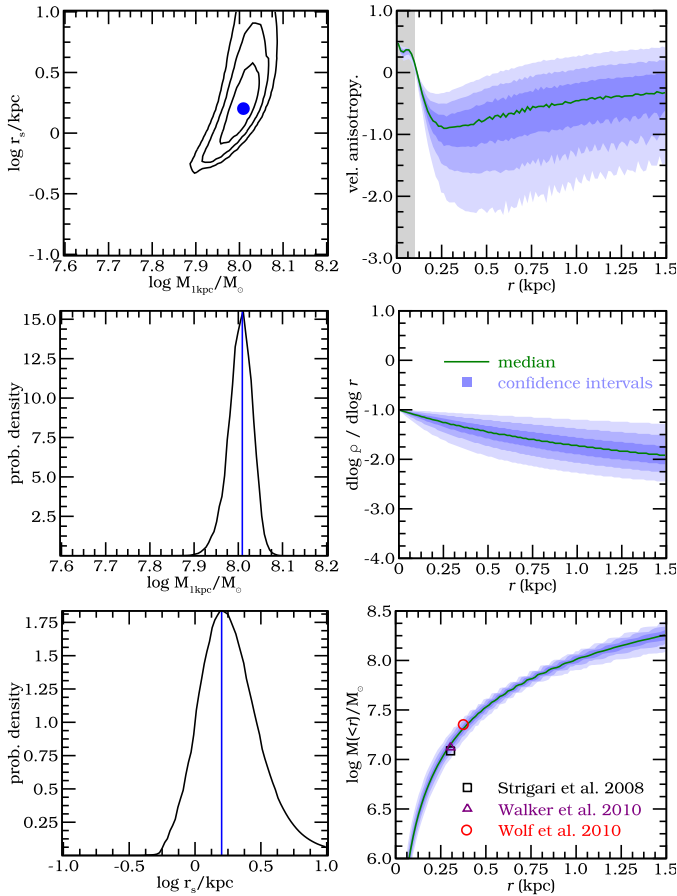}}
\caption{{\bf Left column:} Probability density functions (joint and marginalised) for mass and scale parameters of the NFW dark matter halo potential recovered for Sculptor. Blue dot and blue lines (left column) indicate the maximum likelihood value (of the unmarginalised pdf). The green solid line indicates the median value and the blue regions (or black contour lines in the top left panel) the 68.3, 95.4 and 99.7 per cent confidence intervals. {\bf Top right:} Recovered anisotropy profile. The grey region indicates where we cannot recover the anisotropy. {\bf Middle right:} Recovered logarithmic density slope (see text) for the dark matter. {\bf Bottom right:} Recovered enclosed mass profile. \label{fig:scl-pdf-mass-scale-and-others}}
\end{figure}

We now apply the Schwarzschild method to the data from the Sculptor dSph and model this galaxy as a (non-rotating) spherically symmetric system. For the light distribution we assume a Plummer profile with scale radius $b=0.3$ kpc (Battaglia 2007).

We first assume that Sculptor is embedded in an NFW dark matter halo, as we did for mock Sculptor in \S \ref{sec:testing:nfw}. The results of this modelling are shown in Fig.~\ref{fig:scl-pdf-mass-scale-and-others}. We obtain a tight constraint on the enclosed dark matter mass  of $\Mkpc = 1.03 \times 10^{8.00\pm0.03} \Msol$ (7\% uncertainty, or $\Mkpc = 1.03^{+0.075}_{-0.070}\times10^8  \Msol$). The scale radius at $r_s=2.15\times10^{\pm0.25}$ kpc  (76\% uncertainty, or $r_s=2.15^{+1.6}_{-0.93}$ kpc) is less well constrained, similar to what we find for mock Sculptor.  In comparison to our mock model, the Sculptor dwarf galaxy would seem to have a larger scale radius (see Fig.~\ref{fig:mockscl-pdf-mass-scale-and-others}).

Our estimates are consistent with those derived in previous work for the NFW family of mass models. For example,
\citet{Walker2009ApJ...704.1274W,Walker2010ApJ...710..886W} derive a mass of $10^{+3.2}_{-5.0} \times 10^7 \Msol$ within 1.1 kpc, while
we estimate $10^{+1.3}_{-1.2} \times 10^7\Msol$ within the same distance with smaller error bars. On the other hand, \citet{Battaglia2008ApJ...681L..13B} obtained a mass of $2.2^{+1.0}_{-0.7}\times10^8 \Msol$ within 1.8 kpc, while our measurement at this radius is $1.9^{+0.4}_{-0.3} \times 10^8\Msol$. The mass estimates by \citet{Strigari2008Natur.454.1096S} \citet[][MCMC value]{Walker2010ApJ...710..886W} and \citet{Wolf2010MNRAS.406.1220W} are over plotted in the bottom right panel of Fig.~\ref{fig:scl-pdf-mass-scale-and-others}, and all three agree very well with ours and are within the confidence regions.

The top right panel of  Fig.~\ref{fig:scl-pdf-mass-scale-and-others} shows that Sculptor's anisotropy is mostly tangential and fairly constant with radius, except near the centre where it becomes slightly more isotropic (even after talking into account our limitations due to the projection effects shown and discussed in the context of Fig.~\ref{fig:mockscl-momentfitted-results}). This anisotropy profile at $r>0.1$ kpc is consistent with the constant anisotropy assumed in Jeans models of Sculptor, as by \citet{Walker2007ApJ...667L..53W}, who find $\beta=-0.5$.
\begin{figure}
\centerline{\includegraphics[scale=0.55]{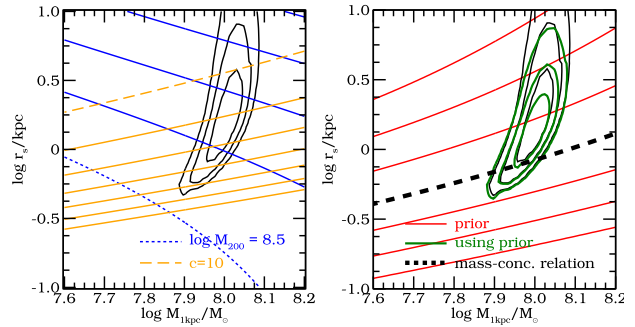}}
\caption{{\bf Left:} The black contours correspond to the same pdf as that shown in the bottom right panel of Fig.~\ref{fig:scl-pdf-mass-scale-and-others}. Blue lines indicate curves of constant $\rm{M}_{200}$, with the blue dotted line corresponding to a value of $\log \rm{M}_{200} = 8.5$, increasing with steps of $0.5$~dex until $\log \rm{M}_{200} = 10.5$. Orange lines indicate values of constant concentration, with the orange dashed line corresponding to $c = 10$, increasing with steps of $5$ until $c = 40$.  {\bf Right:} Red contour lines indicate the cosmologically motivated prior, with the black dashed line the mean value. The green contours are the pdf obtained using this prior for Sculptor.\label{fig:scl-mass-concentration}}
\end{figure}

We plot the joint pdf of $\Mkpc$ and $r_s$ again in Fig.~\ref{fig:scl-mass-concentration}. In the left panel we plot lines of constant virial mass $M_{200}$ in blue\footnote{$M_{200}$ is the virial mass (mass enclosed within $r_{200}$), where $r_{200}$ is the distance at which the average density of a dark matter halo is 200 times the cosmological density $\rho_c$ \citep[e.g.][\S 2.2]{BT}.}, with the blue dotted line indicating a value of $\log \rm{M}_{200} = 8.5$, increasing with steps of $0.5$ dex until $\log \rm{M}_{200} = 10.5$. Orange lines indicate constant concentration values, with the orange dashed line corresponding to $c = 10$, increasing with steps of $5$ until $c = 40$. This shows that the
concentration of Sculptor is $\sim 15 \pm 6$ and that the
 virial mass is not well determined (not better than within factor of 100 at a 3$\sigma$ level uncertainty).

Cosmological N-body simulations of dark matter have shown that there is a relation between the concentration of dark matter halos and their virial masses, the so called mass-concentration relation \citep[e.g.][]{Bullock2001MNRAS.321..559B}. In the right panel of the Fig.~\ref{fig:scl-mass-concentration}, we show as the dashed black line the mass-concentration relation of \citet{Maccio2007MNRAS.378...55M}:
\begin{figure*}
\centerline{\includegraphics[scale=0.9]{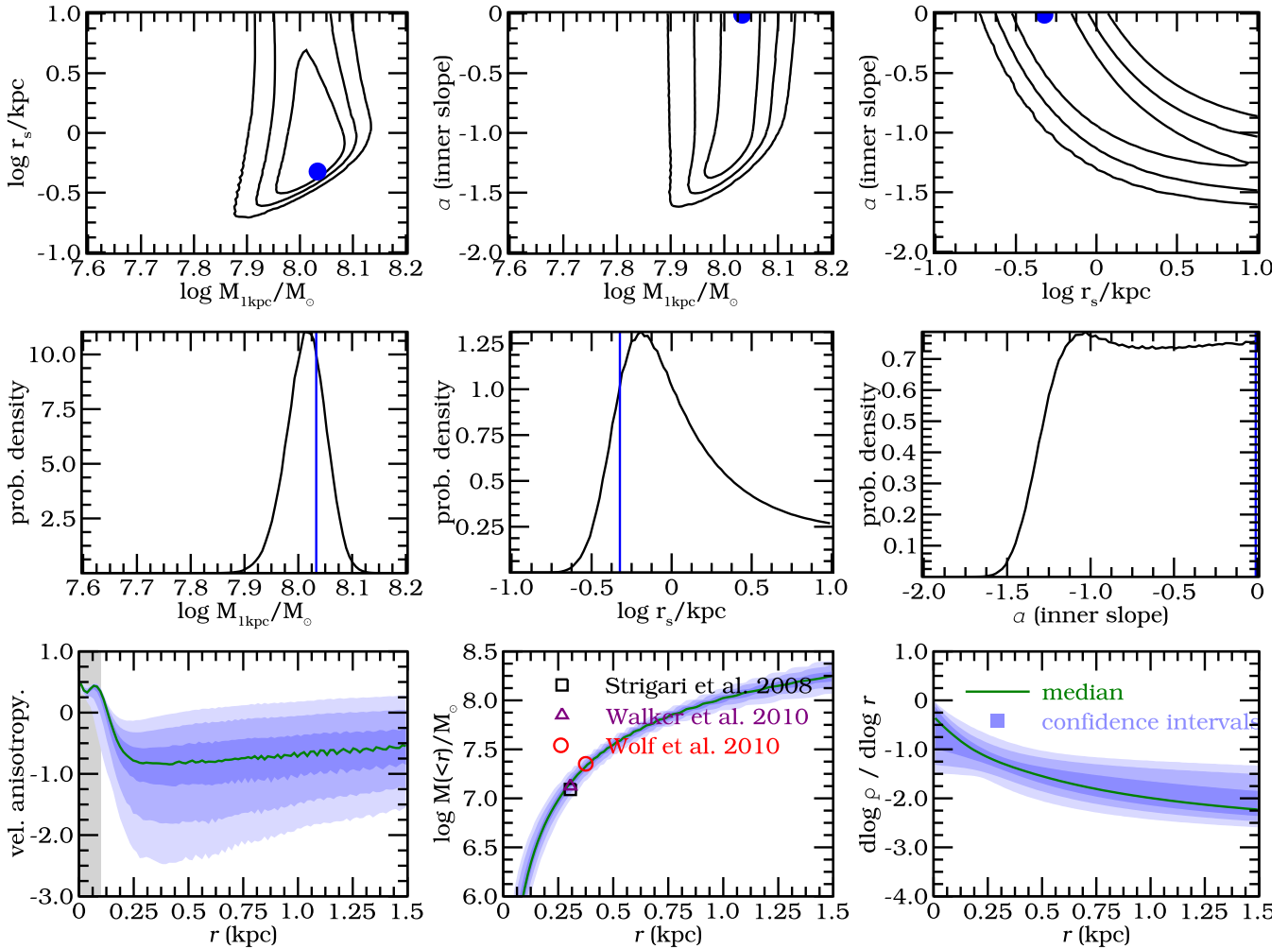}}
\caption{ {\bf Top two rows:} Probability density functions (joint and marginalised) for mass, scale and inner slope parameters of the dark matter halo potential recovered for Sculptor. Blue dots (top row) and blue lines (middle row) indicate the maximum likelihood value (of the unmarginalised pdf). The green solid line indicates the median value and the blue regions (or black contour lines in the top row) the 68.3, 95.4 and 99.7 per cent confidence intervals. {\bf Bottom left:} Recovered anisotropy profile. The grey region indicates where we cannot recover the anisotropy. {\bf Bottom centre:} Recovered enclosed mass profile. {\bf Bottom right:} Recovered logarithmic density slope (see text) for the dark matter.
\label{fig:scl-pdf-mass-scale-slope-and-others}}
\end{figure*}

\begin{equation}
 \log c_{200} = -0.109\log (M_{200}/\Msol) +2.34.
\label{eq:concentr}
\end{equation}
Judging solely from this relationship this would suggest that Sculptor is not compatible with the current $\Lambda$CDM cosmology. If we however plot the intrinsic scatter of $\sigma_{\ln c_{200}} = 0.33$ in the same panel (solid red lines, 1,2 and 3$\sigma$ contours) we see that Sculptor lies well within the 1 and 2$\sigma$ contours. We can also use the mass-concentration relation as a prior in our models. The results are shown as the green contours in this figure and they are slightly smaller than the original contours. The effect is small, but leads to a narrowing down of the possible values for $r_s$.

\subsection{Dark matter inner density profile}

We now consider a more general dark matter profile for the dark matter halo of Sculptor as we did for our mock models in \S \ref{sec:testing:general_halo} by allowing in the inner slope $\alpha$ to vary (see Eq. \ref{eq:rho_general}). The results are shown in Fig.~\ref{fig:scl-pdf-mass-scale-slope-and-others}.

This figure shows that the maximum likelihood value for
$\Mkpc$ and that the velocity anisotropy recovered by the Schwarzschild method are in very good agreement with the values obtained when $\alpha$ is fixed to $-1$ as in Fig.~\ref{fig:scl-pdf-mass-scale-and-others}. However, as discussed in Sec.~\ref{sec:testing:general_halo} the strong degeneracy between $r_s$ and $\alpha$ implies that the scale radius is less well determined.

The middle right panel of Fig.~\ref{fig:scl-pdf-mass-scale-slope-and-others} shows that the distribution of values for the inner slope $\alpha$ is very broad. Nonetheless it is clear that very steep cuspy profiles ($\alpha < -1.5$) are excluded.  The maximum likelihood value is reached for a cored profile ($\alpha=0$), although this is statistically indistinguishable from slightly cuspier slopes as evidenced by the pdfs in this Figure. The bottom right panel of  Fig.~\ref{fig:scl-pdf-mass-scale-slope-and-others} shows that at a distance of 250 pc (where the anisotropy profile begins to change its shape, and which according to our tests in Sec.~\ref{sec:testing:general_halo} is the inner most point where it is reliably determined) the median logarithmic slope profile (green line) takes a value of $\sim -1.25$, which is larger that found in our mock Sculptor model ($\sim -1.75$). Since the maximum likelihood value of $r_s$ estimated by the Schwarzschild method is not very different from that assumed in mock 
Sculptor, this comparison would suggest that the density profile of Sculptor is shallower than NFW, although the uncertainties are still too large to make a very firm statement.

\section{Discussion}
\label{sec:discussion}
\begin{figure*}
\centerline{\includegraphics[scale=0.6]{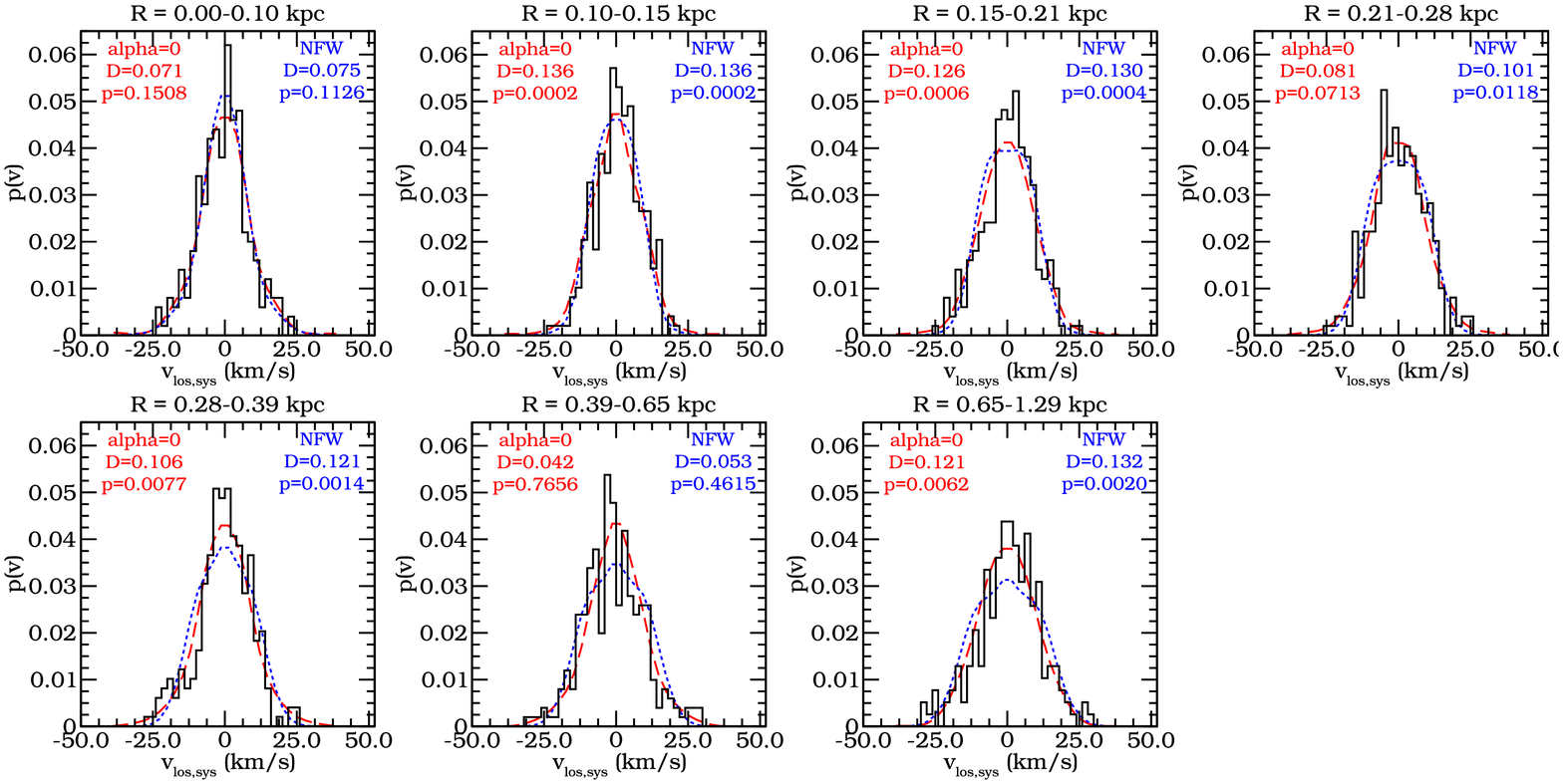}}
\caption{Line-of-sight velocity distributions for Sculptor stars for different radial bins (black histogram). The red curve corresponds to the best-fit {$\alpha = 0$} model, while the best-fit NFW model ($\alpha = -1$) is shown in blue. The $p$-values correspond to the probability that the observed and best-fit model are drawn from the same parent distribution, as quantfied by a KS-test.}\label{fig:histograms}
\end{figure*}

Our results agree with previous studies of Sculptor that {a central logarithmic slope $\alpha = 0$} is
more likely than {the NFW $\alpha = -1$} cusp
\citep{Battaglia2008ApJ...681L..13B,Walker2011ApJ...742...20W,Amorisco2011,Agnello},
although in our case the evidence is clearly not strong enough to rule out the
latter. {Note however, that many, though not all, of these works have tested the presence of a true core, namely $d\rho/dr = 0$ and not just $d \log \rho/dr = \alpha = 0$ at the centre}. A comparison of the statistical significance of our results
with \citet{Walker2011ApJ...742...20W} or \citet{Agnello} is not
straightforward because of the very different methods employed to estimate
the inner slope. These authors use
the existence of two distinct populations (metal-rich and
metal-poor) to constrain the mass distribution (which is modelled non-parametrically), and this may or may not be the cause of the
difference. \citet{Amorisco2011}
favour a cored profile over an NFW with a high significance, but their conclusion is based
on the assumption that two populations follow Michie-King phase-space distribution functions, which are radially anisotropic.
In \citet{boeboe} these authors present evidence that the
velocity anisotropy of Sculptor might in fact radial.  This is in conflict with our results, since we find, with high confidence levels,
that the orbits of stars in Sculptor are tangentially biased (also when marginalised over all models),
especially at radii beyond 250 pc, where the dominant population is
the metal-poor one. Furthermore, also
\citet{Walker2007ApJ...667L..53W, Battaglia2008ApJ...681L..13B} and
\citet{Lokas2009MNRAS.394L.102L} favour a tangentially biased constant anisotropy profile in their Jeans models
of this system.

Given these seemingly contradictory results, it is worthwhile taking a
closer look at the line-of-sight velocity distributions to understand
where the discrepancies might arise. Figure \ref{fig:histograms} shows
these distributions (black histograms) together with the results
obtained for the best fit NFW (blue dotted) and $\alpha = 0$ (red dashed)
models. This figure shows that the l.o.s. velocity distribution is
more peaked at small radii than in the outskirts, where it is more
flat-topped. This is consistent with our measurements of the
l.o.s. kurtosis, and also with our derived anisotropy profile.  As shown by
\citet{Dejonghe1987}, systems with a tangentially biased velocity
ellipsoid have a flat-topped l.o.s. velocity distribution only at
large radii, while in the centre, this distribution is always more
peaked. This is because the l.o.s. towards the centre has
contributions from stars located at a larger range of radii, and hence also from radial
plunging orbits, which drives the shape of the projected velocity distribution to be more
peaked. This is known as the ``complementarity property'', and the
results of our modeling would be consistent with such a scenario.

Figure \ref{fig:histograms}  also shows the small differences between the $\alpha = 0$ and $\alpha = -1$ profiles, and
lend support to our conclusion that the two profiles are both relatively good representations of the data.
This is quantified by a KS-test, whose probabilities are indicated in the corners of each of the panels
of this figure. In a few of the radial bins, none of the models fair particularly well. The $\alpha = 0$ model tends to fit better the
peak of the histogram, and this could be partly to the lower but still tangential anisotropy since $\beta \sim -0.3$ for most radii.

The question thus arises as why do \citet{boeboe} find a radial
anisotropy. Just like us, these authors have utilized the W09
dataset. However, they use stars with a membership probability of 0.5
(as estimated by W09), and do not model the contamination by the Milky
Way any further. In the presence of contaminants, l.o.s. velocity
distributions have extended wings, and this produces a peaky
distribution akin that of truly radially anisotropic systems. We have
tested this idea by measuring the kurtosis for two different
membership probability values $p=0.5$, as in \citet{boeboe}, and
$p=0.9$ (which is more in line with our more sophisticated modeling of
the foreground), and found a significant difference: the kurtosis is
$> 3$ in the first case while in the second case it is consistent with that shown in
Fig. \ref{fig:scl-vlos} of this paper.

In conclusion, care is required when contamination is present, and the differences between {profiles that
have $\alpha = -1$ such as the NFW or $\alpha = 0$}, although present, are perhaps not as dramatic as maintained in other published work.

\section{Conclusions}
\label{sec:conclusions}

We have presented a spherically symmetric dynamical model for the Sculptor dwarf spheroidal galaxy using the Schwarzschild orbit superposition method. This method fits a set of observables, which in our case are the light, the second and fourth moments of the line of sight velocity distribution. We have tested this method on a mock model for the Sculptor dSph galaxy embedded in an NFW profile, and generated with similar sampling and velocity errors as the data currently available for this system.

In our tests we have found our method to give precise ($7\%$ uncertainty) and accurate estimates for the mass within 1 kpc, when assuming that the underlying gravitational potential is of NFW form. However the scale radius is recovered less precisely  ($37\%$ uncertainty)  for data sets containing $\sim 2000$ member stars. We have also explored a more general model for the dark matter halo and found that we are able to measure the logarithmic slope of its density profile, although the central value is weakly constrained. Nonetheless we find that the maximum likelihood value for the inner slope is very close to the input value.

We then used the Schwarzschild method on Sculptor after having estimated the second and fourth line of sight velocity moments for this galaxy. Assuming an NFW profile for the dark matter profile, we derive a mass within 1 kpc of $\Mkpc = (1.03 \pm 0.07 ) \times 10^{8}$ \Msol, and find
the concentration ($c \sim 15$) to be compatible with current $\Lambda$CDM predictions, given the expected scatter in the mass-concentration relation \citep{Maccio2007MNRAS.378...55M}. When we try to constrain the inner slope of the dark matter density profile of Sculptor, we can exclude very cuspy profiles ($\alpha < -1.5$). However, given the current data set, our method does not seem to be able to discriminate in a statistically significant way between {a $\alpha = -1$ cusp and a central logarithmic slope $\alpha = 0$}, although the latter is the most likely value. We are, however, able to determine that the logarithmic slope of the density profile falls off to the value of $-2$ at a distance of $\sim 1$ kpc.

The Schwarzschild method is also able to derive the velocity anisotropy profile, except near the centre where we are limited by the number of tracers. For Sculptor we find this to be tangentially biased with a hint that it may become more isotropic for $r \lesssim 250$ pc. This result is nearly independent of the assumed shape of the dark matter density profile, whether NFW or its generalised form. This nearly flat tangentially anisotropic ellipsoid should hold clues to the formation and dynamical evolution of Sculptor but it is as yet unclear whether a model exists that can reproduce this trend.

Models in which stars follow the dark matter are inconsistent with our results, as they predict a more radially anisotropic velocity ellipsoid \citep{Diemand2004MNRAS.352..535D}. On the other hand, the tidal stirring of a disky galaxy \citep[see e.g.][]{Mayer2010AdAst2010E..25M}, can lead to a tangentially biased ellipsoid. However, this model predicts that the ellipsoid becomes increasingly tangential with radius as a consequence also of tidal stripping, and this is not what we derive at face value.

Schwarzschild modelling does not have to assume a parametric form for the velocity anisotropy as for instance in the commonly used Jeans modelling. We therefore believe that we are less affected by biases due to assumptions compared to such class of models. Furthermore, by construction we are guaranteed that our models are physical in the sense of having non-negative distribution functions.

We plan to develop the Schwarzschild method further to work with the full line of sight velocity distribution, instead of binning the data and comparing it the the velocity moments profile. Avoiding the loss of information when binning, we expect that this may give us better estimates for the inner slope and the anisotropy profile. Also, since neither Sculptor nor any of the other dwarf spheroidal galaxies are spherical, we are developing a non spherical orbit-based dynamical model. We also plan to apply this modelling to other dwarf spheroidal galaxies such as Fornax, Carina and Sextans in future work.

\section*{Acknowledgements}
AH and MB are grateful to NOVA for financial support. AH acknowledges financial support the European Research Council under ERC-Starting Grant
GALACTICA-240271. We are also grateful to Carlos Vera-Ciro and Simon White for various interesting discussions.

\bibliographystyle{mn2e}
\bibliography{schwscl}

\appendix

\section{Numerical approximation to the distribution function}
\label{sec:df}

We take the following separable form for the distribution function:
\begin{eqnarray}
 f(E, L) &=& f_{E}(E)f_L(L).
\end{eqnarray}
For a constant anisotropy for instance, $f_L(L) \propto L^{-2\beta}$. Now we assume that the distribution function $f(E,L)$ can be approximated by $\hat{f}(E,L)$, where $\hat{f}_{E}(E)$ is a sum of delta functions, such that:
\begin{eqnarray}
 \hat{f}(E, L) &=& \frac{1}{N}\sum_{i=1}^{N} w_i \delta(E-E_i)f_L(L) \label{eq:df:sum}.
\end{eqnarray}
The density distribution corresponding to this distribution function is:
\begin{eqnarray}
 \hat{\nu}(r) &=& 2\pi \int_{-v_{r,\text{max}}}^{v_{r,\text{max}}} dv_r \int_0^{v_{t,\text{max}}} v_t dv_t \hat{f}(E,L)\\
	&=& \frac{4\pi}{r^2} \int_0^{-\Phi(r)} \!\!\!\!\!\!dE \int_0^{L_\text{max}} \!\!\!\!\!\!dL L\frac{ \hat{f}(E,L), }{\sqrt{-2(E-\Phi(r))-\frac{L^2}{r^2}}}\\
	&=& \frac{4\pi}{r^2} \frac{1}{N}\sum_{i=1}^{N} w_i \int_0^{L_\text{max}} \!\!\!\!\!\!dL L\frac{ f_L(L)}{\sqrt{-2(E_i-\Phi(r))-\frac{L^2}{r^2}}}\\
	& &  \times \Theta(-(E_i-\Phi(r)))\\
	&=& \frac{1}{N}\sum_{i=1}^{N} w_i \hat{\nu}_{i}(r),
\end{eqnarray}
where $\Theta$ is the Heaviside step function and the $\hat{\nu}_{i}(r)$ the densities that correspond to the each of the energy delta functions.

Given a stellar density distribution $\nu(r)$ and a gravitational potential $\Phi(r)$, it may be possible to find the weights $w_i$ such that $\nu(r) \approx \hat{\nu}(r)$. In this case we may state that we have found a numerical approximation to the distribution function that generates the proper stellar density distribution and is embedded in the potential $\Phi(r)$. A solution can be found for instance using a non-negative least square method. An even simpler method is to start with the $\hat{\nu}_{j}$ corresponding to the lowest binding energy. All $\hat{\nu}_{i}$ associated with higher binding energies can only contribute to the density at smaller radii, therefore by weighing $\hat{\nu}_{j}$ this can account for the density out to  the outermost radius. Now one can proceed with the next $\hat{\nu}_{i}$. Thus we start from the lowest binding energy components, use appropriate weights and build the density distribution from outside in. Care should be taken to make sure all weights are positive. 

In the case of the mock Sculptor model discussed in the main text, $\nu(r)$ is the Plummer profile and $\Phi(r)$ is the sum of the potentials of the Plummer mass distribution describing the stellar component and that generated by the NFW profile associated to the dark halo. In this case we have chosen $f_L(L) \propto L^{-2\beta}$, where $\beta = -1$. For our purpose we choose a logarithmically spaced radial grid of 600 points between $r_\text{min}=10^{-3}$ kpc and $r_\text{max}=10^{3}$ kpc. For each $r_i$ on the grid, we calculate the potential energy, giving us a grid of energies, which we take the energies for our distribution function ($E_i$ in Eq.~\ref{eq:df:sum}). For each $E_i$ we calculate the density on the same radial grid. The last step is to find the weights $w_i$ using the above procedure. A small mismatch (few \%) of the density at large radii ($>300$ pc) occurs due to the distribution function missing lower binding energy components. The cumulative mass distribution of the stellar mass deviates 
$<10^{-4}$ from the true mass distribution, and within $300$ pc the relative density deviates $<2\times10^{-4}$. Outside this radius the density does not match very well, but since this is at large radii and its mass contribution is very small (note also that the cumulative mass distribution shows only small deviations) this is of no importance.

\section{Centre of mass velocity of Sculptor}
\label{sec:app:center_of_mass}

In this Appendix we transform the observed line-of-sight velocities to velocities with respect to
the centre of mass of Sculptor. This requires knowledge of the latter, which is what we derive here using
a maximum likelihood method.

The observed (heliocentric) line-of-sight velocity of a star can be expressed as:
\begin{equation}\begin{split}
 v_{*,hel}(l, b) &= \myvec{e}_{los}(l, b) \boldsymbol{\cdot} \left( \myvec{v}_{*,Scl}(l, b) + \myvec{v}_{Scl,GSR} - \myvec{v}_{\odot, GSR} \right)\\
 & = v_{*,Scl}(l, b) + v_{Scl,GSR}(l, b) - v_{\odot, GSR}(l, b),\end{split}\nonumber\end{equation}
where $\myvec{e}_{los}(l, b)$ is the line-of-sight unit vector in the direction of the star, $\myvec{v}_{*,Scl}(l, b)$ the velocity of the star with respect to the centre of mass of Sculptor, $\myvec{v}_{Scl,GSR}$ the systemic velocity of the centre of mass of Sculptor with respect to the Galactic Standard of Rest (hereafter GSR), $v_{\odot, GSR}$ the velocity of the Sun with respect to the GSR and $\boldsymbol{\cdot}$ indicates the inner product. The component of the line-of-sight velocity we are interested in is $v_{*,Scl}(l, b)$. Since $v_{*,hel}(l, b)$ is measured, and assuming we know $\myvec{v}_{\odot, GSR}$, we only need to find $v_{Scl,GSR}(l, b)$. For the velocity of the Sun we use $v_{\odot, GSR} = v_{\odot, LSR} + v_{LSR, GSR} = (10.0, 5.2, 7.2) + (0, 220, 0) ~\mathrm{\kms}$, where $_{LSR}$ denotes Local Standard of Rest \citep{DehnenBinney1998}.

\begin{figure}
\centerline{\includegraphics[scale=0.45]{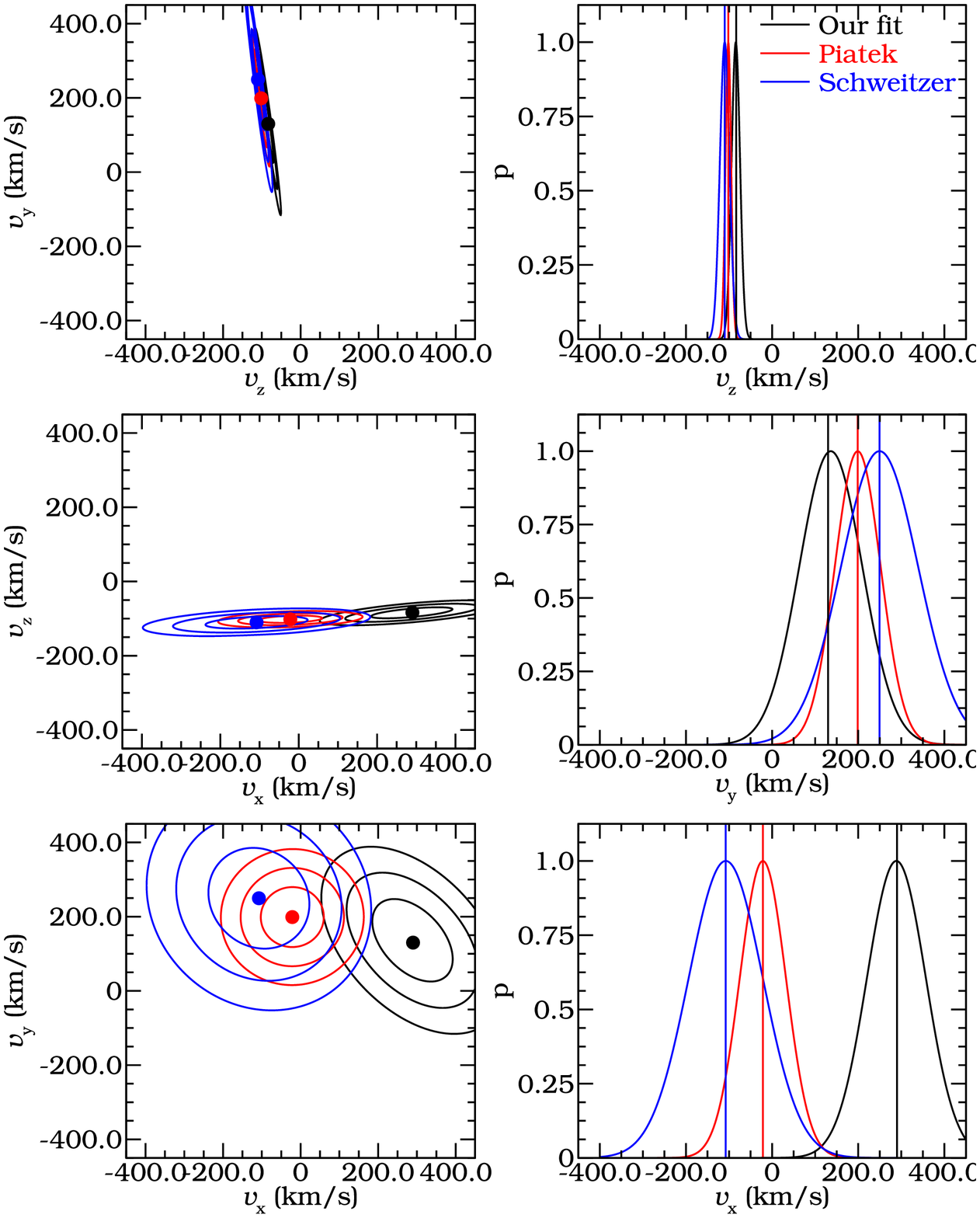}}
\caption{Probability distribution function (pdf) of the three velocity components of the systemic velocity of Sculptor with respect to the Galactic Standard of Rest. {\bf Left column:} Joint pdfs with 1, 2 and 3 $\sigma$ contours lines, marginalised over the other component. {\bf Right column:} Individual pdfs marginalised over the other two components. The measurements from \citet{Piatek2006AJ....131.1445P} are shown in red, while those by \citet{Schweitzer1995AJ....110.2747S} are shown in blue. The vertical lines in the right panels and the dot in the left panels indicate the maximum likelihood values. \label{fig:scl-v-gsr}}
\end{figure}

To determine which stars are likely members of Sculptor we make a rough first selection. We take the systemic heliocentric radial velocity ($v_\text{Scl,sys,helio} = 110.6$ km s$^{-1}$)  and the mean velocity dispersion ($\sigma_\text{Scl} = 10.1$ km s$^{-1}$) from \citet{Battaglia2008ApJ...681L..13B}. We first require that the member stars are within $3\sigma$ of the systemic velocity of Sculptor, as indicated by the red solid lines in the right panel of Fig. \ref{fig:scl-vlos}. Furthermore we also require that they are located within $r~<~0.944$ degree, indicated by the green dashed line in the same panel. We add this requirement since we are not confident that outside this radius a reliable velocity dispersion can be measured due to the low number density of (probable) Sculptor members compared to Milky Way stars.

For simplicity we first assume that the line-of-sight velocity distribution is described by a Gaussian distribution with a constant velocity dispersion and zero mean velocity w.r.t the centre of mass of Sculptor. Then the probability for $\myvec{v}_{Scl,GSR}$ can be expressed as:
\begin{equation}\begin{split}
 p(\myvec{v}_{Scl,GSR}) &= \prod_i \frac{1}{\sqrt{2\pi} \sigma_i} \exp{\left( -\frac{v_{*_i,Scl}(l_i, b_i)^2}{2\sigma_i^2}\right) }\\
	&= \prod_i \frac{1}{\sqrt{2\pi} \sigma_i} \exp{ \left[ -\frac{1}{2\sigma_i^2} \left\{ v_{*_i,hel} - \right. \right. } \\
&  \left. \vphantom{\frac{1}{2\sigma_i^2}} \left.  \myvec{e}_{los}(l_i, b_i) \cdot \left(\myvec{v}_{\odot, GSR}-\myvec{v}_{Scl,GSR} \right) \right\}^2 \right]\end{split}
\end{equation}
where $\sigma_i^2 = \sigma_{Scl}^2 + \sigma_{*_i}^2$ is the velocity dispersion of Sculptor added in quadrature with the measurement error of the velocity of star $i$. Although the velocity dispersion is not constant with radius, we use the global value of $\sigma_\text{Scl} = 10.1~\kms$ as described previously.

The joint and marginalised probability distribution functions for the velocity components of Sculptor are plotted in Fig. \ref{fig:scl-v-gsr} together with the 1, 2 and 3 $\sigma$ contours. The maximum likelihood value is reached at $\hat{\myvec{v}}_{Scl,GSR} = (v_x, v_y, v_z) = (278.5, 101.5, -81.0)~\kms$. These values are in agreement with \citet{Walker2008ApJ}, who use a similar method. We also over plot the measurements of \citet[][in red]{Piatek2006AJ....131.1445P}  and  \citet[][in blue]{Schweitzer1995AJ....110.2747S} while the maximum likelihood value is indicated in black. Note that the uncertainty in $v_z$ is smallest since this reflects mainly the uncertainty in the mean radial velocity of the centre of mass of Sculptor due to its high galactic latitude. The uncertainties in the other two velocity components mainly reflect the uncertainties in
the proper motion measurements. Our determination of the $v_y$ component agrees well with the various data sets, while the $v_x$ component appears to be systematically offset. Note however, that there is overlap at the 3$\sigma$ level, and the $2\sigma$ and $3\sigma$ contours for the joint $v_x$ and $v_y$  overlap as well. Perhaps this level of disagreement could be taken as an indication that there may be intrinsic rotation in the system. Nonetheless, we note that with this procedure we effectively have removed the observed gradient and no apparent rotation remains, whatever its origin.

\end{document}